\documentclass[12pt]{article}
\usepackage{savesym,dsfont,ytableau}
\usepackage{cite}
\usepackage{picture}
\usepackage{wasysym}
\usepackage{tikz, tikz-cd}
\usepackage{amscd}
\usepackage{graphicx}
\usepackage{pifont}
\usepackage{float}
\usepackage{subfig}
\usepackage[all]{xy}
\usepackage{multicol}
\usepackage{amsfonts}
\usepackage{picture}
\usepackage{amssymb}
\savesymbol{iint}
\savesymbol{iiint}
\usepackage{amsmath}
\restoresymbol{TXF}{iint}
\restoresymbol{TXF}{iiint}
\usepackage{color}
\usepackage{clay}
\usepackage{hyperref}
\usepackage{slashed}
\hypersetup{colorlinks=true}
\hypersetup{linkcolor=black}
\hypersetup{citecolor=black}
\hypersetup{urlcolor=black}
\usepackage{setspace}
\usepackage{multirow}
\usepackage{wrapfig}
\usepackage{verbatim}
\usepackage{accents}
\numberwithin{equation}{section}

\newcommand\fro{{\overline{\underline{\Omega}}}}

\newcommand*{\dt}[1]{
  \accentset{\mbox{\large.}}{#1}}
\def\dot{\dt}

\begin{document}

\date{March, 2017}

\institution{IAS}{\centerline{${}^{1}$School of Natural Sciences, Institute for Advanced Study, Princeton, NJ, USA}}
\institution{PI}{\centerline{${}^{2}$Perimeter Institute for Theoretical Physics, Waterloo, Ontario, Canada N2L 2Y5}}

\title{Surface Defect Indices and 2d-4d BPS States}

\authors{Clay C\'{o}rdova\worksat{\IAS}\footnote{e-mail: {\tt clay.cordova@gmail.com}}, Davide Gaiotto\worksat{\PI}\footnote{e-mail: {\tt dgaiotto@perimeterinstitute.ca}},  and Shu-Heng Shao\worksat{\IAS}\footnote{e-mail: {\tt shuhengshao@gmail.com}} }

\abstract{We conjecture a formula for the Schur index of four-dimensional $\mathcal{N}=2$ theories coupled to $(2,2)$ surface defects in terms of the $2d$-$4d$ BPS spectrum in the Coulomb phase of the theory.  The key ingredient in our conjecture is a refined $2d$-$4d$ wall-crossing invariant, which we also formulate.  Our result intertwines recent conjectures expressing the four-dimensional Schur index in terms of infrared BPS particles, with the Cecotti-Vafa formula for limits of the elliptic genus in terms of two-dimensional BPS solitons.  We extend our discussion to framed $2d$-$4d$ BPS states, and use this to demonstrate a general relationship between surface defect indices and line defect indices.  We illustrate our results in the example of $\frak{su}(2)$ super Yang-Mills coupled to the $\mathbb{CP}^1$ sigma model defect.}

\maketitle
{\setstretch{1.05}
\setcounter{tocdepth}{3}
\tableofcontents
}
\section{Introduction}

In this paper we discuss four-dimensional $\mathcal{N}=2$ theories coupled to two-dimensional $(2,2)$ surface defects $\mathbb{S}$.  We investigate the defect Schur index
\begin{equation}
\mathcal{I}_{\mathbb{S}}(q)=\sum_{\mathcal{O}_{2d-4d}}\left[e^{2\pi iR}q^{R-M_{\perp}}\right]~,\label{SSchurdefintro}
\end{equation}
where in the above $R$ denotes the  $\frak{su}(2)$ $R$-charge and $M_{\perp}$ rotations transverse to the defect.  This index counts supersymmetric local operators bound to the defect.  

We state a refined $2d$-$4d$ wall-crossing formula which governs the discontinuities in the BPS particle and soliton spectrum of the defect $\mathbb{S}$, and conjecture a formula for $\mathcal{I}_{\mathbb{S}}(q)$ using the particle spectrum.   The $2d$-$4d$ solitons and particles entering the conjecture are defined in the Coulomb phase of the theory and as such our conjecture may be viewed as an infrared formula for the defect Schur index.  

\subsection{Surface Defects}

Surface defects are objects supported along two-dimensional manifolds in spacetime.  Like their one-dimensional cousins, i.e. line defects, these objects can be useful probes for investigating the phases of gauge theories or exploring non-perturbative phenomena such as dualities.  In the context of supersymmetric four-dimensional theories, there are supersymmetric surface defects whose properties may be explored explicitly.  Our focus in this paper is on those defects that preserve $(2,2)$ supersymmetry.  A review of many of the properties of these defects is given in \cite{Gukov:2014gja}.  

There are a variety of defects that may be considered.  One may introduce fields living on a two-dimensional locus, and couple them to the bulk degrees of freedom.  A prototypical example that has been widely investigated is the two-dimensional $\mathbb{CP}^{N-1}$ sigma-model coupled to four-dimensional $\frak{su}(N)$ Super Yang-Mills theory by gauging the global symmetry of the defect \cite{Gaiotto:2013sma}.  Other examples are disorder type defects where the bulk fields are singular along a two-dimensional manifold \cite{Gukov:2006jk, Gukov:2008sn} (the holographic description of disorder defects was given in \cite{Gomis:2007fi}). In certain cases, these two constructions of defects may be related by dualities.  Another realization of many surface defects arises from the construction of $\mathcal{N}=2$ theories in terms of M5-branes \cite{Hanany:1997vm, Gaiotto:2009fs, Gaiotto:2011tf}.  

In the presence of a surface defect $\mathbb{S},$ there is a rich set of physical questions that may be investigated. Our main aim is to elucidate a connection between two conceptually distinct ideas that may be viewed as ultraviolet and infrared data of the theory in the presence of the defect.
\begin{itemize}
\item UV data:  In the ultraviolet the defect $\mathbb{S}$ may be characterized by the spectrum of local operators that are bound to the defect.  A partial count of these operators is given by the defect generalization of the superconformal index.  For Lagrangian examples this may be computed using supersymmetric localization \cite{Nakayama:2011pa, Gadde:2013ftv}.  Its general properties have been explored in \cite{Gaiotto:2012xa, Alday:2013kda, Bullimore:2014nla}.

\item IR data:  The defect may be investigated by moving onto the Coulomb branch and flowing to the infrared, generalizing the $4d$ bulk dynamics of \cite{Seiberg:1994rs, Seiberg:1994aj}.  In the examples considered here the defect degrees of freedom are gapped in the IR.  In each $2d$ vacuum one finds a collection of  $2d$ BPS particles.  Additionally, one finds a set of BPS solitons that interpolate between distinct vacua.  These states may carry four-dimensional electromagnetic charges and are referred to as $2d$-$4d$ BPS states \cite{Gaiotto:2011tf}. Aspects of the $2d$-$4d$ BPS spectrum have appeared in \cite{Longhi:2012mj, Gaiotto:2012rg, Longhi:2016rjt, Longhi:2016bte}.  As moduli are varied the $2d$-$4d$ BPS spectrum may jump according to a wall-crossing formula. 
\end{itemize}

These two disparate ideas of the IR BPS particle spectrum, and the UV local operator spectrum are linked by our results.  Indeed we will present a limit of the defect index as a wall-crossing invariant generating function of $2d$-$4d$ BPS particles.  In the special case where the bulk theory is empty our results reduce to the well-known Cecotti-Vafa formulas \cite{Cecotti:1992rm} for limits of the elliptic genus in terms of the $2d$ BPS spectrum.  In the special case where the surface defect is empty our results reduce to the formulas of \cite{Cordova:2015nma, Cecotti:2015lab, Cordova:2016uwk} expressing the ordinary Schur index in terms of bulk BPS particles.  The general case we present is a hybrid of these two formalisms.

\subsection{Schur Indices}\label{sec:schur}

The limit of the index that we reproduce using BPS states is the defect generalization of the Schur index $\mathcal{I}(q)$ introduced in \cite{Kinney:2005ej, Gadde:2011ik, Gadde:2011uv}.  The Schur index depends on a single universal fugacity $q,$ and links a variety of topics in mathematical physics, including topological field theory \cite{Gadde:2009kb, Gadde:2011ik}, vertex operator algebras \cite{Beem:2013sza}, and BPS wall-crossing phenomena \cite{Iqbal:2012xm, Cordova:2015nma}.

It is useful to view the Schur index in greater detail to understand exactly why it appears naturally in our UV/IR relation between local operators and particles.  Examining the definition \eqref{SSchurdefintro}, we see a sum over local operators weighted by their $\frak{su}(2)_{R}$ charge and spins.  Note that the scaling dimension and $\frak{u}(1)_{r}$ charges do not appear.  Thus it is natural to expect that the symmetries associated to these quantum numbers are not needed to define the index.  This idea has been made precise in \cite{DFZ} by defining the Schur index for non-conformal $\mathcal{N}=2$ theories as a partition function on the $S^{3}\times S^{1}$.  In particular, the Coulomb branch theory, including its massive BPS particle excitations, has a well-defined Schur index.  

When the bulk theory is coupled to a surface defect $\mathbb{S}$, there exists an appropriate generalization of the Schur index that we describe in detail in section \ref{sec:indicesgen}.  In the special case where the $4d$ dynamics are trivial, this index reduces to the limit of the $2d$ NS-NS sector (2,2) elliptic genus which counts operators that are simultaneously chiral with respect to both the left and right supersymmetry algebras.  More generally, one can view the defect Schur index $\mathcal{I}_{\mathbb{S}}(q)$ as counting simultaneously chiral operators bound to the defect.  

A crucial feature of $2d$ $(2,2)$ surface defects in $4d$ $\mathcal{N}=2$ theories is that they always possess a flavor symmetry $\frak{u}(1)_{C},$ which arises from the superalgebra embedding.  The  generator $C$ of this flavor symmetry descends from the bulk charges as:
\begin{equation}
C=R-M_{\perp}~.
\end{equation}
Thus, from the $2d$ point of view, the universal parameter $q$ appearing in the defect Schur index is a flavor fugacity which further grades the chiral operator spectrum.

A powerful perspective on the Schur index $\mathcal{I}(q)$ was introduced in \cite{Beem:2013sza} and further developed in \cite{Beem:2014rza, Lemos:2014lua, Liendo:2015ofa, Lemos:2015orc, arakawa2015joseph, Nishinaka:2016hbw, Buican:2016arp, Arakawa:2016hkg, Bonetti:2016nma, beem}.  There it was argued on general grounds that for a conformal field theory, the local operators contributing to $\mathcal{I}(q)$ form a two-dimensional non-unitary chiral algebra.  As a consequence, the Schur index is the vacuum character of this chiral algebra.  This is a strong organizing principle and has been utilized to great effect to understand aspects of the operator algebra of $\mathcal{N}=2$ SCFTs.  The chiral algebra aspects of surface defects will be explored in \cite{CGS, BPR}.

\subsection{The Refined $2d$-$4d$ Wall-Crossing Formula}

The main ingredient in our infrared formulas for surface defect Schur indices is the refined $2d$-$4d$ BPS spectrum and wall-crossing formula.  As compared to \cite{Gaiotto:2011tf}, which studied the unrefined indices and wall-crossing formula, the $2d$-$4d$ BPS counts appearing in our work are graded by the universal flavor charge $C$.  Note that $C$ includes transverse rotations to the defect so effectively we are refining the BPS degeneracies by their four-dimensional spin.

We assemble a wall-crossing invariant spectrum generator $\mathcal{S}^{2d-4d}_{\vartheta, \vartheta'}(X_{\gamma})$ from these refined degeneracies generalizing the ideas of \cite{Kontsevich:2008fj, Gaiotto:2008cd, Dimofte:2009tm, Dimofte:2009bv, Gaiotto:2009hg, Gaiotto:2012rg, Galakhov:2014xba} for pure $4d$ systems as well as the extensions to $4d$ systems coupled to defects \cite{Gaiotto:2010be,  Gaiotto:2011tf}.  The object $\mathcal{S}^{2d-4d}_{\vartheta, \vartheta'}(X_{\gamma})$ is an $N\times N$ matrix, with $N$ the number of vacua of the defect $\mathbb{S}$, whose entries are valued in a quantum torus algebra constructed from variables $\{X_{\gamma}\}$ where $\gamma$ is any elctromagnetic charge.  The variables obey
\begin{equation}
X_{\gamma}X_{\gamma'}=q^{\frac{1}{2}\langle\gamma, \gamma'\rangle}X_{\gamma+\gamma'}~, \label{qtintro}
\end{equation}
with $\langle\gamma, \gamma'\rangle$ the Dirac pairing.  

The spectrum generator $\mathcal{S}^{2d-4d}_{\vartheta, \vartheta'}(X_{\gamma})$ encodes the part of the spectrum whose central charge phases lie in the wedge $\vartheta \leq \arg(\mathcal{Z})\leq \vartheta'$.  It is written as a phase ordered product of factors, where each factor is associated to either a $4d$ bulk BPS particle ($K^{4d}$) or a $2d$ BPS particle  ($K^{2d}$) or soliton interpolating between the $i$-th and $j$-th vacuum ($S_{ij}$).  Thus:
\begin{equation}\label{refinedwall2dintro}
\mathcal{S}^{2d-4d}_{\vartheta_1,\vartheta_2}(q)= : \prod_{\{ij,\gamma| \arg(\mathcal{Z})\in [\vartheta_1 , \vartheta_2)\}}^{\curvearrowleft} 
S_{ij}(X_{\gamma})K^{2d}(X_{\gamma}) K^{4d}(X_{\gamma}) :
~,
\end{equation}
where the normal ordering notation above indicates that the various factor matrices should be ordered according to the phase of their central charge.  As parameters are varied, the spectrum, and hence the decomposition of the spectrum generator into factors, jumps.  However the product is invariant.  We present this formalism in detail and give simple examples of wall-crossing in section \ref{sec:2d4dBPS}.

\subsection{The Infrared Formula for Surface Defect Indices}\label{sec:IRintro}

It is straightforward to anticipate the form of our conjecture for the defect Schur index given the previous results on infrared formulas for Schur indices without surface defects \cite{Cordova:2015nma, Cordova:2016uwk} as well as the Cecotti-Vafa formula \cite{Cecotti:1992rm} which express limits of the elliptic genus in terms of $2d$ solitons.

The key idea pioneered in \cite{Cecotti:1992rm, Cecotti:2010fi, Iqbal:2012xm} is to extract functions from wall-crossing invariant operators generating the BPS spectrum.  In the context of surface defects, the natural object is the $2d$-$4d$ spectrum generator $\mathcal{S}^{2d-4d}$ introduced above, and our conjectured formula for the surface defect Schur index in terms of the infrared BPS spectrum reads:
\begin{equation}
\mathcal{I}_{\mathbb{S}}(q)=(q)_{\infty}^{2r}~\mathrm{Tr}\left[\phantom{\int}\hspace{-.17in}\mathcal{S}^{2d-4d}_{\vartheta, \vartheta+\pi}(X_{\gamma})\mathcal{S}^{2d-4d}_{\vartheta+\pi, \vartheta+2\pi}(X_{\gamma})\right]~. \label{conjintro}
\end{equation}
Here, $r$ is the rank of the Coulomb branch, and the trace operation appearing above is an ordinary matrix trace on the matrix degrees of freedom, as well as a trace on the quantum torus algebra.

This conjecture, as well as its simpler versions without surface defects \cite{Cordova:2015nma, Cordova:2016uwk}, admits a simple heuristic understanding:  it is the index as computed using the IR description as an abelian gauge theory, where we simply proceed as if the individual BPS particles arose from independent free fields.  This is of course a naive idea: the true IR effective field theory on the Coulomb branch cannot be described so simply, but this perspective is nevertheless useful for understanding the structure of \eqref{conjintro}.  Indeed, the factor $(q)_{\infty}^{2}$ is the index of a free abelian vector multiplet.  Meanwhile the contributions of the $4d$ BPS particles are simply the indices of hypermultiplets (appropriately modified to include spin).  The trace then selects from the product those states which carry vanishing electric and magnetic charges.  

From this intuitive point of view, the appearance of the quantum torus algebra \eqref{qtintro} has a simple physical interpretation: it is modifying the quantum numbers of ``composite operators."  To elaborate on this, imagine independent hypermultiplet fields $H_{\gamma}$ carrying electromagnetic charge $\gamma$ and spin $J_{\gamma}$.  We must then confront the question of what quantum numbers to assign to the composite
\begin{equation}
H_{\gamma}(0)H_{\gamma'}(0)~.
\end{equation}
The quantum torus algebra appearing in our infrared conjectures tells us that this operator should carry electromagnetic charge $\gamma+\gamma'$, but spin $J_{\gamma}+J_{\gamma'}+\frac{1}{2}\langle \gamma, \gamma'\rangle.$  The shift proportional to the Dirac pairing accounts for the angular momentum in the induced electromagnetic field.  Of course, this introduces an ordering ambiguity, which is resolved by the normal ordering prescription introduced above.

\subsection{Relations Between Line Defects and Surface Defects}

A conceptual implication of our infrared formalism for defect Schur indices is a general relationship between line defects and surface defects that we develop in section \ref{sec:linesurrel}.  

\begin{figure}[h!]
\begin{center}
\includegraphics[width=.5\textwidth]{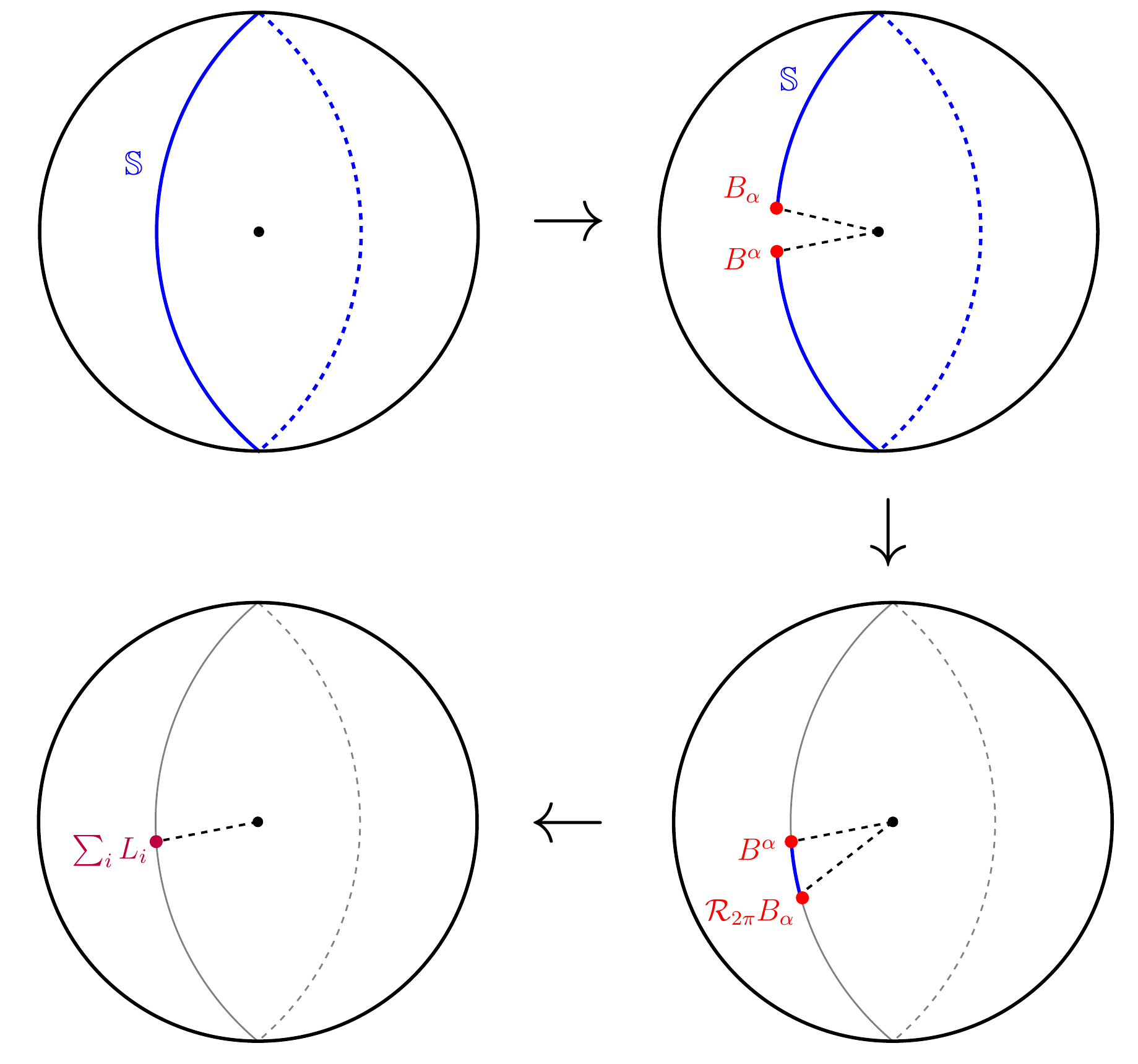}
\end{center}
\caption{ The unwrapping process of a surface defect to a sum of lines.  Starting in the upper left we have a surface  defect $\mathbb{S}$ (shown in blue) wrapping an equator in $S^{3}$ as well as the $S^{1}$ which is not shown.  The defect is cut open using a resolution of the identity with left and  right boundary conditions $B^{\alpha}$ and $B_{\alpha}$.  The boundary conditions are then parallel transported around the equator using the operator $\mathcal{R}_{2\pi}$.  When they collide the surface defect is gone and a sum of line defects $L_{i}$ remains.}\label{fig:cut}
\end{figure}

Given a surface defect $\mathbb{S}$ it is frequently possible to resolve the identity interface on $\mathbb{S}$ into a sum of left and right boundary conditions.  If we insert this resolution in the $S^{3}\times S^{1}$ index geometry we can then unwrap the surface defect into a sum of lines $L_{i}$ as illustrated in Figure \ref{fig:cut}.  The purely two-dimensional version of this cutting procedure was discussed in \cite{Gaiotto:2015zna, Gaiotto:2015aoa}, while aspects of the extension to $2d$-$4d$ coupled systems were described in \cite{Gaiotto:2011tf}. 

This geometric manipulation implies a general relationship between the surface defect index $\mathcal{I}_{\mathbb{S}}(q)$ and the line defect indices $\mathcal{I}_{L}(q):$ 
\begin{equation}
\mathcal{I}_{\mathbb{S}}(q) =\sum_{j}c_{j}(q)\mathcal{I}_{L_{j}}(q)~, \label{linesurfrelintro}
\end{equation}
where the $c_{j}(q)$ are coefficients defined in the unwrapping process.  Moreover using the infrared formula for line defect indices \cite{Cordova:2016uwk } (reviewed in section \ref{sec:linedef}), together with technology for computing framed BPS degeneracies \cite{Gaiotto:2010be, Lee:2011ph, Cordova:2013bza, Galakhov:2014xba, Moore:2015szp, Gabella:2016zxu, Brennan:2016znk}, as well as some technology for dealing with framed $2d$-$4d$ degeneracies developed in sections \ref{pure2dframed} and \ref{sec:linesurrel}, these decompositions may be determined explicitly.

To motivate these results we recall that in \cite{Cecotti:2010fi, Cordova:2016uwk} it was noticed that Schur indices in the presence of BPS line defects frequently give rise to sums of characters of chiral algebras. These observations are mysterious, and beg interpretation.  The relation between line defect indices and surface defect indices to a large extent explains these observations, and will be discussed in more detail in \cite{CGS}.

\subsection{$\frak{su}(2)$ Super Yang-Mills Coupled to the $\mathbb{CP}^{1}$ Sigma Model}

Finally, in section \ref{sec:simex} we provide a non-trivial check on our formalism in the example of $\frak{su}(2)$ super Yang-Mills coupled to its canonical surface defect, the $\mathbb{CP}^{1}$ sigma model.

In this case, since both the defect and the bulk have Lagrangian descriptions, the defect Schur index can be computed using localization techniques resulting in the formula
\begin{equation}
\mathcal{I}_{\mathbb{S}}(q) =   \oint \frac{du}{2\pi i u} \frac{(1-u^{2})(1-u^{-2})}{2}(qu^{2};q)_{\infty}^{2}(q;q)_{\infty}^{2}(qu^{-2};q)_{\infty}^{2}(-u^{2}-u^{-2}) ~.
\end{equation}
The integral can be done explicitly resulting in a Jacobi theta function, $\theta_{3}(2\tau)$.

We compare this result to the prediction from our infrared formula \eqref{conjintro} using the $2d$-$4d$ BPS spectrum determined in \cite{Gaiotto:2011tf}.  This results in
\begin{equation}\label{SU(2)IRintro}
 \mathcal{I}_\mathbb{S}(q)=(q)_\infty^2 \sum_{\ell_1,\ell_2=0}^\infty {q^{\ell_1+\ell_2+2\ell_1\ell_2} \over (q)_{\ell_1}^2 (q)_{\ell_2}^2} \left(
2q^{-\ell_2} - q^{\ell_1-\ell_2} -q^{-\ell_1-\ell_2} +2q^{-\ell_1} -q^{-\ell_1+\ell_2}
\right)~.
\end{equation}
Strikingly we find a perfect match between these two expressions. 

We also discuss resolving the $\mathbb{CP}^{1}$ surface defects into line defects and establish a relationship: 
\begin{equation}
\mathcal{I}_{\mathbb{S}}(q)=\mathcal{I}(q)-\mathcal{I}_{L_{\mathbf{3}}}(q)~,
\end{equation}
where on the right-hand side above, $\mathcal{I}(q)$ is the ordinary Schur index, and $\mathcal{I}_{L_{\mathbf{3}}}(q)$ is the index in the presence of an adjoint Wilson line.
 
In \cite{CGS} we investigate more examples of the IR formula and give further evidence for its validity.

\section{$2d$-$4d$ Coupled Systems and Their Indices}
\label{sec:indicesgen}

In this section we discuss basic properties of $2d$-$4d$ coupled systems and their indices.  In section \ref{sec:surfdefstuff} we describe basic properties of surface defects.  In section \ref{sec:2d4dindicesbasic} we discuss $2d$-$4d$ Schur indices.

\subsection{Surface Defects in $\mathcal{N}$=2 Theories}
\label{sec:surfdefstuff}

Given a four-dimensional $\mathcal{N}=2$ theory, we may couple the bulk fields to degrees of freedom that reside on a locus of codimension two in spacetime.  This construction defines a surface defect, $\mathbb{S}$, of the bulk theory.  We are interested in surface defects that preserve some of the initial supersymmetry of the bulk system.  The bulk system may or may not enjoy additional conformal supersymmetry, and the defect can preserve or break the scale invariance of the bulk SCFT. 

The most constrained possibility is that of a $4d$ SCFT coupled to a conformally invariant surface defect.  The bulk superconformal algebra is $\frak{su}(2,2|2).$  Its bosonic subalgebra is
\begin{equation}
\frak{so}(4,2)\times \frak{su}(2)_{R} \times \frak{u}(1)_{r}\subset \frak{su}(2,2|2)~,
\end{equation} 
consisting of the conformal algebra as well as $R$-symmetries.  We consider half-BPS surface defects which preserve two-dimensional $(2,2)$ supersymmetry.\footnote{A second class of half-BPS defects preserving (0,4) supersymmetry also exists.  They are not discussed in this paper.}  These defects are extended along a plane in spacetime.  They preserve the superalgebra
\begin{equation}
 \frak{su}(1,1|1)\times  \frak{su}(1,1|1)\times \frak{u}(1)_{C}\subset \frak{su}(2,2|2)~. \label{conf2d}
\end{equation}
The $\frak{su}(1,1|1)$ factors yield the global part of the $2d$ (2,2) superconformal algebra while $\frak{u}(1)_{C}$ is the commutant of the embedding.  The symmetry $\frak{u}(1)_{C}$ is therefore a universal flavor symmetry of $2d$ (2,2) surface defects and will play a crucial role throughout the paper.

It is instructive to match the various Cartan generators in the $2d$ defect superalgebra to their $4d$ bulk origin.  Let $\Delta, R, r$ denote the four-dimensional scaling charge, $\frak{su}(2)_{R}$ Cartan, and $\frak{u}(1)_{r}$ charge respectively, and let $M_{||}$ and $M_{\perp}$ denote the rotations along the defect plane and orthogonal to the defect plane.  Then, as reviewed in appendix \ref{chargedetails}, the relationship of these generators to the $2d$ chiral scaling generators $L_{0}, \bar{L}_{0},$ and $R$-charges $J_{0}, \bar{J}_{0}$ is
\begin{equation}
L_{0}=\frac{1}{2}(\Delta+M_{||})~,\hspace{.4in} \bar{L}_{0}=\frac{1}{2}(\Delta-M_{||})~,\hspace{.4in} J_{0}=2R-M_{\perp}+r~,\hspace{.4in}\bar J_0 = - 2R + M_{\perp}+r~. \label{cartanrels}
\end{equation}
Meanwhile the universal flavor symmetry $C$ descends from the bulk algebra as
\begin{equation}
C=R-M_{\perp}~.\label{cartanrels1}
\end{equation}

More generally, we may obtain the superalgebra for non-conformal defects by restricting both the bulk and defect superalgebras to the superPoincar\'{e} generators.  Note that in such a reduction the universal flavor symmetry survives as long as the bulk theory preserves $\frak{su}(2)_{R}$.

It is significant that the charges appearing in \eqref{conf2d} are the global part of the super Virasoro algebra which reside in the NS-NS sector.  In $2d$ (2,2) SCFTs with compact spectrum, there is a Virasoro enhancement of the charge algebra and also Ramond sectors.  By contrast $2d$ defects do not in general exhibit such enhancement, and their worldvolume typically does not support a conserved energy-momentum tensor.  Thus, the charges that are manifest in \eqref{conf2d} are all that are generally present in coupled $2d$-$4d$ systems.

Given a $4d$ theory, there are a wide variety of possible surface defects $\mathbb{S}$ to consider.  A common construction is to restrict the four-dimensional bulk fields to the two-dimensional defect locus and then couple them to fields residing on the defect.  The bulk $4d$ vector multiplet restricts to a $2d$ vector multiplet, or equivalently, a twisted chiral multiplet.  We may utilize this to make non-trivial defects in Lagrangian field theories as follows.
\begin{itemize}
\item The $4d$ vector multiplet may participate in a twisted superpotential $\widetilde{W}$.
\item The $4d$ vector multiplet may gauge a $2d$ flavor symmetry.
\end{itemize}
Alternative constructions of surface defects appear in non-Lagrangian theories.  For instance, in theories of class $S$ obtained by compactification of the six-dimensional $(2,0)$ theory on a Riemann surface $\Sigma$ a set of defects called \emph{canonical surface defects} arise from M2-branes residing at a point on $\Sigma$ \cite{Gaiotto:2009fs,Gaiotto:2011tf}.  

\subsection{$2d$-$4d$ Indices and the Schur Limit}
\label{sec:2d4dindicesbasic}

Given a $2d$-$4d$ coupled system we are interested in the spectrum of local operators bound to the surface defect $\mathbb{S}$.  We obtain information about this sector of operators by computing the superconformal index.  In this section we review the basic features of this index and the Schur limit.  

\subsubsection{Schur Indices in $4d$ Field Theories}

We begin with a $4d$ $\mathcal{N}=2$ SCFT.  The general superconformal index is \cite{Kinney:2005ej,Gadde:2011uv}\footnote{We continue to use the notation $M_{||}$ and $M_{\perp}$ for the rotation Cartans in order to match onto our conventions for surface defects.  The rotation generators of the two independent $\frak{su}(2)$'s in the Lorentz group are expressed as
\begin{equation}
M_{||}=j_{2}-j_{1}~,\hspace{.5in}M_{\perp}=-(j_{1}+j_{2})~.
\end{equation}
} 
\begin{equation}
\mathcal{I}(q,p,t,x)=\mathrm{Tr}\left[\phantom{\int}\hspace{-.13in}e^{2\pi i R}~ q^{-M_{\perp}-r}p^{M_{||}-r}t^{R+r}x^{H}\right]~. \label{Igendef}
\end{equation}
Here the trace is over the Hilbert space on $S^{3}$, or equivalently by the state operator correspondence, a sum over local operators.  

The variables $p,q, t$ are universal while the $x$ variables account for possible flavor charges $H$ in the theory.  By construction the operators that contribute to the index are at least $1/8$-BPS (annihilated by two odd generators of the superconformal algebra) and satisfy
\begin{equation}
 \Delta +M_{\perp}-M_{||}-2R+r=0~.
\end{equation}
The index also vanishes on all combinations of short multiplets that may recombine into long multiplets and hence is stable under marginal deformations.  In particular this implies that $\mathcal{I}(q,p,t,x)$ may be easily computed for theories with a Lagrangian definition by working at zero coupling.

In the remainder of this work, we are interested in the Schur limit of the superconformal index \cite{Gadde:2011ik,Gadde:2011uv}.  This is a limit of the general index where only $1/4$-BPS operators contribute to the sum.  These operators satisfy the two shortening conditions
\begin{equation}
 \Delta +M_{\perp}-2R=0~, \hspace{.5in}r-M_{||}=0~. \label{schurshort}
\end{equation}
The Schur index depends on a single universal fugacity $q$ and flavor variables $x$. It is obtained from the general definition \eqref{Igendef} by specializing $t\rightarrow q.$ The additional shortening conditions \eqref{schurshort} then imply that the result does not depend on the variable $p$.  This index may be expressed as  a sum over local operators of the form 
\begin{equation}
\mathcal{I}(q,x)=\sum_{\mathcal{O}}\left[e^{2\pi iR}q^{R-M_{\perp}}x^{H}\right]~.\label{Schurdef}
\end{equation}
As with the full index we may alternatively view $\mathcal{I}(q,x)$ as a trace over the Hilbert space on $S^{3}$.  

Note that in both \eqref{Igendef} and \eqref{Schurdef} we have used a slightly unconventional fermion number
\begin{equation}\label{4dfermion}
F_{4d}=2R~,
\end{equation}
as opposed to $F_{4d}=2(j_{1}+j_{2})$.  As a consequence of the shortening condition \eqref{schurshort} the two conventions are related by modifying the Schur index as $q^{1/2}\rightarrow -q^{1/2}$.

One interesting feature of the Schur index presented as in \eqref{Schurdef} is that it does not refer to either the scaling dimension $\Delta$ or the $\frak{u}(1)_{r}$ charge $r$.  Therefore it is natural to expect that the existence of these symmetries is not necessary to define the Schur index.  Instead, we can take the operator sum \eqref{Schurdef} as a definition of the Schur index.  In particular, this definition may be applied to non-conformal $\mathcal{N}=2$ theories (both asymptotically free and infrared free) provided only that they have an $\frak{su}(2)_{R}$ $R$-symmetry.  We will see examples of such indices for non-conformal theories in later sections.  The Schur index for non-conformal $\mathcal{N}=2$ theories also admits a definition in terms of a supersymmetric path integral on $S^{3}\times S^{1}$ \cite{DFZ}.  

For Lagrangian field theories, the Schur index may be computed by enumerating gauge invariant operators built of free fields.  Since the answer is an index, the result is not modified by the interactions.  We can use this to obtain a simple matrix integral expression for the Schur index.  We require the single letter partition functions for vector multiplets and hypermultiplets\footnote{The unconventional sign in $f^{\frac 12 H}$ comes from our choice of the 4$d$ fermion number \eqref{4dfermion}.}
\begin{equation}
f^{V}(q)=-\frac{2q}{1-q}~, \hspace{.5in} f^{\frac{1}{2}H}=-\frac{q^{1/2}}{1-q}~,
\end{equation}
which determine the contribution to the index from the field operators.  The contributions of products of these operators are taken into account using the plethyestic exponential 
\begin{equation}
P.E.[f(q,u,x)]=\exp\left[\sum_{n=1}^{\infty}\frac{1}{n}f(q^{n},u^n,x^{n})\right]~.
\end{equation}
If we introduce Pochhammer symbols defined as
\begin{equation}
(a;q)_n  
=\begin{cases}  1 & n=0~,\\ \prod_{j=0}^{n-1}(1-a q^{j}) & n>0~,\end{cases}
\end{equation}
and set $(q)_n \equiv (q;q)_n$, then we can also write the plethyestic exponential as
\begin{equation}\label{fVH}
P.E.[f^V(q) u]=(q u;q)_\infty^2 ~, \hspace{.5in} P.E.[f^{\frac{1}{2}H}(q)u]=(-q^{1/2} u;q)_\infty^{-1} ~.
\end{equation}

For a theory with gauge group $G$ and hypermultiplet fields in a representations $\mathbf{R}$ of $G$ and $\mathbf{F}$ of the flavor symmetry we then have 
\begin{equation}
\mathcal{I}(q,x)=\int[du] \,P.E. \left[f^{V}(q)\chi_{G}(u)+f^{\frac{1}{2}H}(q)\chi_{\mathbf{R}}(u)\chi_{\mathbf{F}}(x)\right]~,\label{operatorSchur}
\end{equation}
where $\chi_{\alpha}$ are characters of the gauge and flavor group and $[du]$ is the Haar measure on the maximal torus of $G$.\footnote{A helpful identity in explicit calculations is to rewrite the half-hypermultiplet contribution as 
\begin{equation}
(q^{\frac12} x;q)^{-1}_\infty = \sum_{n=0}^{\infty}\frac{(q^{\frac{1}{2}}x)^{n}}{(q)_{n}}~.
\end{equation}
The integral over the gauge fugacities can then be done explicitly. }

\paragraph{Example: $\frak{su}(2)$ SYM}\mbox{}\\
An explicit example that will be useful in later sections is the Schur index of $\frak{su}(2)$ super Yang-Mill theory.  Note that this is not a conformal field theory, but we may still compute it's Schur index using the operator sum definition.  We obtain
\begin{equation}
\mathcal{I}(q)=\oint \frac{du}{2\pi i u} \frac{(1-u^{2})(1-u^{-2})}{2}(qu^{2};q)_{\infty}^{2}(q;q)_{\infty}^{2}(qu^{-2};q)_{\infty}^{2}=\sum_{n=0}^{\infty}q^{n(n+1)}=\frac{q^{-1/8}}{2}\theta_{2}(2\tau)~, \label{su2schur}
\end{equation}
where in the above we have used the Jacobi triple product identity.  

\subsubsection{Elliptic Genera in $2d$ Field Theories}

Let us now turn to indices in $2d$ theories.  These are elliptic genera.  In applications to  $2d$ conformal field theories it is common to consider the elliptic genus in the Ramond-Ramond sector.  This is natural when the elliptic genus is viewed as a partition function on the torus.  However, in this work we will instead formulate the elliptic genus as an operator counting generating function.  Such a perspective is more natural in the Neveu-Schwarz-Neveu-Schwarz sector.  Therefore we formulate the genus as a trace over the radially quantized Hilbert space in this sector. This NS-NS sector genus is also compatible with the coupled $2d$-$4d$ systems discussed in the previous section.  Indeed in that context, the only supercharges that are generally available on a $2d$ defect reside in the Neveu-Schwarz-Neveu-Schwarz sector.

The explicit formula for the genus of interest takes the form:\footnote{The variable $\mathbf{q}$ that appears in the following is unrelated to the variable $q$ appearing in the Schur index.}  
\begin{equation}
\mathcal{G}(\mathbf{q},y,u)=\mathrm{Tr}_{NSNS}\left[(-1)^{F_{2d}}\mathbf{q}^{L_0}\bar{\mathbf{q}}^{\bar L_0-\bar J_0/2}y^{J_{0}}u^{K}\right]~. \label{NSNSEG}
\end{equation}
In the above,  $L_0$ and $\bar L_0$ are the left and right Hamiltonians, $J_0$ and $\bar J_{0}$ are the left and right $\frak{u}(1)$ $R$-symmetries, and $K$ are flavor charges with associated variables $u$. Finally, $F_{2d}$ is the fermion number defined in terms of the left and right $R$-symmetry charges as
\begin{equation}\label{2dfermion}
F_{2d}=J_0+\bar J_0~.
\end{equation}
Note that as a consequence of this definition, the fermion number $F_{2d}$ may be an arbitrary real number.  Due to supersymmetry, the elliptic genus is independent of $\bar{\mathbf{q}}$ and receives contributions only from states on the right which are chiral primaries and hence satisfy
\begin{equation}
\bar L_0-\bar J_0/2=0~.\label{antichiral}
\end{equation}

Although the NS-NS sector genus is all that we will encounter in later sections, it is instructive to consider the relationship to the more standard Ramond sector genus,
\begin{equation}
\mathcal{G}_{RR}(\mathbf{q},y,u)=\mathrm{Tr}_{RR}\left[(-1)^{F_{2d}}\mathbf{q}^{L_0-c/24}\bar{\mathbf{q}}^{\bar L_0-c/24}y^{J_{0}}u^{K}\right]~, \label{RREG}
\end{equation}
where in the above,  $c=c_{L}=c_{R}$ is the central charge. Due to supersymmetry, the RR sector elliptic genus is also independent of $\bar{\mathbf{q}}$ and receives contributions only from right moving Ramond sector ground states.  

In a general two-dimensional CFT, spectral flow symmetry relates the two elliptic genera.  Let $(h,\rho)$ denote a weight and $R$-symmetry charge (either left or right).  Then flow by $\eta$ units acts on $(h,\rho)$ as
\begin{equation}
(h,\rho)\mapsto (h+\eta \rho+\eta^{2}c/6, \rho+\eta c/3)~.
\end{equation}
By taking $\eta=1/2$ both on left and on the right we transition from the RR to the NSNS sector and hence deduce that the genera are related as
\begin{equation}
\mathcal{G}_{RR}(\mathbf{q},y)=y^{-c/6}\mathcal{G}_{NSNS}(\mathbf{q},\mathbf{q}^{-1/2}y)~. \label{RRNNrel}
\end{equation}

As with the four-dimensional superconformal index, our focus is on a simplified limits of the elliptic genus, which are analogous to the four-dimensional Schur limit.  There are two possible limits of \eqref{NSNSEG} which enjoy enhanced supersymmetry.  These are the specialized genera where $\mathbf{q}=y^{\pm2}$
\begin{equation}
\mathcal{G}_{(c,c)}(u)\equiv\mathcal{G}(\mathbf{q},y,u)|_{\mathbf{q}=y^{-2}}~, \hspace{.5in}\mathcal{G}_{(a,c)}(u)\equiv\mathcal{G}(\mathbf{q},y,u)|_{\mathbf{q}=y^{2}}~. \label{special}
\end{equation}
The specialized index $\mathcal{G}_{(c,c)}(u)$ counts operators which are simultaneously chiral on both the left and on the right, while the specialized index $\mathcal{G}_{(a,c)}(u)$ counts operators which are left antichiral and right chiral.  These indices depend only on flavor parameters $u$. 

The specialized elliptic genera enjoy many parallels with the four-dimensional Schur index $\mathcal{I}(q)$.   Among these, is the fact that they may be defined for non-conformal $(2,2)$ theories where some of the $R$-symmetries are broken.   Indeed, instead of first formulating the full elliptic genus and then taking a limit, we may define the specialized genera by summing over the restricted set of chiral $\times$ chiral or antichrial$\times$chiral operators.  Therefore the specialized genera can be defined as long as these shortening conditions on operators may be formulated.

To clarify this further, let us introduce the Poincar\'{e} supercharges for the $(2,2)$ algebra. 
\begin{equation}
\def\arraystretch{1.5}
\begin{tabular}{c|c|c|c|c|}
& $L_{0}$ & $\bar{L}_{0}$&$J_{0}$&$\bar{J}_{0}$\\
\hline
$Q_{+}$&$1/2$ &$0$&$+1$&$0$\\
\hline
$\bar{Q}_{+}$& $1/2$&$0$&$-1$&$0$\\
\hline
$Q_{-}$& $0$&$1/2$&$0$&$+1$\\
\hline
$\bar{Q}_{-}$&$0$ &$1/2$&$0$&$-1$\\
\hline
\end{tabular}
\label{qcharges}
\end{equation}
To formulate the specialized genera, it is sufficient that various supercharges anticommute so that we may define a simultaneous shortening condition.  Specifically,
\begin{equation}
\mathcal{G}_{(c,c)}(u)~ \mathrm{exists}\Leftrightarrow \{Q_{+},Q_{-}\}=0~, \hspace{.5in}\mathcal{G}_{(a,c)}(u)~ \mathrm{exists}\Leftrightarrow \{Q_{+},\bar{Q}_{-}\}=0~. \label{scftcomms}
\end{equation}
These requirements may be usefully phrased in terms of the vectorial ($R_{V}$) and axial ($R_{A}$) $R$-symmetries defined as
\begin{equation}
R_{V}=J_{0}+\bar{J}_{0}~,\hspace{.5in}R_{A}=J_{0}-\bar{J}_{0}~.
\end{equation}
Away from a fixed point, the superconformal algebra is deformed (see, for example,  \eqref{massivealgebra}) and the right-hand side of the commutators \eqref{scftcomms} may be non-vanishing. Accordingly, some $R$-symmetries are broken.  Depending on which of  $R_V$ and $R_A$ is preserved, one of the two specialized elliptic genera can be defined for the non-conformal theory.  We summarize this discussion as below:
\begin{equation}
R_{V}~ \mathrm{preserved} \Rightarrow \mathcal{G}_{(c,c)}(u)~ \mathrm{exists}~,\hspace{.5in}R_{A}~ \mathrm{preserved} \Rightarrow \mathcal{G}_{(a,c)}(u)~ \mathrm{exists}~.
\end{equation}

Finally, let us also comment on dependence of elliptic genera on flavor variables $u$.  Using the superconformal algebra, it can be shown that in any compact $(2,2)$ all the (anti)chiral primaries are uncharged under any flavor symmetries. This implies that for all compact  SCFTs the specialized index is a number that does not depend on any variables.  By contrast for non-conformal theories, or conformal theories with non-compact spectra the dependence on flavor variables is in general non-trivial.

For theories with a Lagrangian description the elliptic genus may be computed by counting free fields \cite{Gadde:2013ftv,Benini:2013nda,Benini:2013xpa}.  Since the resulting quantity is an index it is invariant upon activating interactions.  In practice in the following we will only use these tools to compute the genera of theories with UV descriptions as Abelian gauge theories coupled to chiral multiplets, possibly with superpotentials. See \cite{Gadde:2013ftv,Benini:2013xpa} for a more complete treatment including the case of non-abelian gauge theories.

To express the formula for the elliptic genus of an abelian gauge theory, we require the theta function
\begin{align}
\theta_1(\mathbf{q},x) = -i \mathbf{q}^{1/8} x^{1/2}\prod_{k=1}^\infty (1-\mathbf{q}^k)(1-x\mathbf{q}^k)(1-x^{-1}\mathbf{q}^{k-1})~.
\end{align}
Then, the elliptic genus of a chiral with left $R$ charge $J_{0}=r$ (and hence $\bar{J}_{0}=r$) is given by
\begin{equation}
\mathcal{G}_{\mathrm{chiral}}(\mathbf{q},y, u)=\mathbf{q}^{1\over4} y^{-{1\over2}} {\theta_1(\mathbf{q},\mathbf{q}^{-{(1+r)\over2}}  y^{1-r} u^{-1}) \over \theta_1(\mathbf{q}, \mathbf{q}^{-\frac{r}{2}}y^{-r} u^{-1}) }~, \label{chiralEG}
\end{equation}
where $u$ is a flavor fugacity for the $\frak{u}(1)$ global symmetry.  Similarly, the elliptic genus of a twisted chiral multiplet is obtained from the above by changing $J_{0}\rightarrow-J_{0}$ and hence changing $y\rightarrow y^{-1}$.

The $(2,2)$ $\frak{u}(1)$ vector multiplet may be viewed as a twisted chiral with $r=1$.  This leads to\footnote{Technically, the twisted chiral elliptic genus evaluated at $r=1$ vanishes due to the existence of a supersymmetry breaking superpotential.  The vector multiplet elliptic genus given below is obtained by removing that zero mode, which is then integrated in the localization formulas that follow.}
\begin{equation}
\mathcal{G}_{\mathrm{vector}}(\mathbf{q},y)=-i\mathbf{q}^{-1/8} y^{1/2}{(\mathbf{q})_\infty^3  \over \theta_1(\mathbf{q},\mathbf{q}^{-{1\over2}} y)}~.
\end{equation}

From these ingredients we can formulate the index of a $\frak{u}(1)$ gauge theory coupled to chiral multipets $\Phi_{i}$.  Let $c_{i}$ denote the gauge charges, $r_{i}$ the $R$-charges, and $f_{i}$ the flavor charges.  Then the resulting genus is
\begin{equation}
\mathcal{G}(\mathbf{q},y, u)= -i\mathbf{q}^{-1/8} y^{1/2}{(\mathbf{q})_\infty^3  \over \theta_1(\mathbf{q},\mathbf{q}^{-{1\over2}} y)}\sum_{z_{j}\in \mathcal{M}_{+}}\oint_{z=z_{j}} \frac{dz}{2\pi i z} \prod_{i}\mathbf{q}^{1\over4} y^{-{1\over2}} {\theta_1(\mathbf{q},\mathbf{q}^{-{(1+r_{i})\over2}}  y^{1-r_{i}} z^{c_{i}}u^{-f_{i}}) \over \theta_1(\mathbf{q}, \mathbf{q}^{-\frac{r_{i}}{2}}y^{-r_{i}} z^{c_{i}}u^{-f_{i}}) }~. \label{local}
\end{equation}
Where in the above the meaning of the sum is as follows.  The integrand has poles where the theta function in the denominator vanishes.  We sum over those poles $z_{j}$ that arise from particles of positive charge $c_{i}>0$.

\paragraph{Example: $(2,2)$ Minimal Models}\mbox{}\\

Consider the $(2,2)$ minimal model described as a Landau-Ginzburg theory of a single twisted chiral field $\widetilde{\Phi}$ with twisted superpotential $\widetilde{W}=\widetilde{\Phi}^{n+1}$.  The right-moving $R$-charge is $r=(n+1)^{-1}$ and hence the elliptic genus of this theory is given by 
\begin{equation}
\mathcal{G}(\mathbf{q},y, u)=\mathbf{q}^{1\over4} y^{{1\over2}} {\theta_1\left(\mathbf{q},\mathbf{q}^{-\frac{n+2}{2(n+1)}}  y^{-\frac{n}{n+1}}\right) \over \theta_1\left(\mathbf{q}, \mathbf{q}^{-\frac{1}{2(n+1)}}y^{\frac{1}{n+1}} \right) }~.
\end{equation}
Upon spectral flow to the RR sector, this reproduces the result of \cite{Witten:1986bf}.

We can extract the specialized indices by taking the limit $y\rightarrow \mathbf{q}^{\pm\frac{1}{2}}$ (taking care that the theta functions may vanish in these limits). We obtain
\begin{equation}
\mathcal{G}_{(c,c)}=1~, \hspace{.5in}\mathcal{G}_{(a,c)}=n~. \label{mmspecialized}
\end{equation}
These are the expected results for the specialized genera.  Indeed the twisted superpotential truncates the twisted chiral ring leaving the $n$ operators $1, \widetilde{\Phi}, \widetilde{\Phi}^{2}, \cdots, \widetilde{\Phi}^{n-1}$ which are counted by $\mathcal{G}_{(a,c)}$. Meanwhile the only $(c,c)$ operator in the theory is the identity.  Note as expected that both specialized general are independent of $\mathbf{q}$.

\paragraph{Example: $\mathbb{CP}^{1}$ Sigma Model}\mbox{}\\

The $\mathbb{CP}^1$ sigma model can be defined in the ultraviolet by a $\frak{u}(1)$ gauge theory with two positively charged chiral multiplets, and a positive FI parameter which sets the size for the $\mathbb{CP}^1$ in the infrared.  In this model, the axial $R$-symmetry $R_{A}$ is anomalous.  The theory is non-conformal and gapped in the infrared.  Correspondingly the integrand appearing in the the localization formula \eqref{local} is not single-valued and the general elliptic genus is ambiguous.  However, since the vectorial $R$ symmetry $R_{V}$ remains a symmetry we may still compute the specialized genus $\mathcal{G}_{(c,c)}$. We obtain the following expression 
\begin{eqnarray}
\mathcal{G}_{(c,c)}(u)&=&\lim_{q\rightarrow y^{-2}}-i\mathbf{q}^{3/8} y^{-1/2} {(\mathbf{q})_\infty^3  \over \theta_1(\mathbf{q},\mathbf{q}^{-{1\over2}} y)}\sum_\pm \oint_{z=\mu^{\pm1}} {dz\over 2\pi iz }
{\theta_1(\mathbf{q},\mathbf{q}^{-{1\over2}}  y zu)\theta_1(\mathbf{q},\mathbf{q}^{-{1\over2}}  y zu^{-1}) \over \theta_1(\mathbf{q},zu)\theta_1(\mathbf{q},z u^{-1}) }\nonumber\\
 &=& \lim_{q\rightarrow y^{-2}}\mathbf{q}^{1/4}y^{-1/2} \left( {\theta_1(\mathbf{q},\mathbf{q}^{-{1\over2}}  y u^{2}) \over \theta_1(\mathbf{q},u^{2}) }+{\theta_1(\mathbf{q},\mathbf{q}^{-{1\over2}}  y u^{-2}) \over \theta_1(\mathbf{q},u^{-2}) }\right) \\ \label{CP1special}
 & = & 1-(1+u^{2}+u^{-2})~,\nonumber
\end{eqnarray}
where $u$ is the fugacity for the $\frak{su}(2)$ flavor symmetry.  Note that the flavor dependence of the specialized genus is non-trivial which is permitted since this example is not a conformal field theory.  The final answer captures the expected $(c,c)$ operators: the identity, and an $\frak{su}(2)$ triplet realized geometrically by holomorphic vector fields.\footnote{Recall that $(c,c)$, or B-model, local operators in non-linear sigma models are associated to the $\bar \partial$ cohomology valued in holomorphic poly-vector fields. See e.g. \cite{Bershadsky:1993cx}.}

\subsubsection{$2d$-$4d$ Indices}\label{sec:2d4dindices}

We now fuse the previous discussions to obtain Schur indices in the presence of surface defects.  Abstractly the indices of interest count local operators in the presence of a surface defect $\mathbb{S}$
\begin{equation}
\mathcal{I}_{\mathbb{S}}(q,x)=\sum_{\mathcal{O}_{2d-4d}}\left[e^{2\pi iR}q^{R-M_{\perp}}x^{H}\right]~.\label{SSchurdef}
\end{equation}
This index may be viewed as a hybrid between the $2d$ specialized elliptic genus $\mathcal{G}_{(c,c)}$ and the $4d$ Schur index.  Indeed, using the relation \eqref{cartanrels} between the $4d$ and $2d$ algebras, we deduce that the $2d$ charges obey
\begin{equation}
2L_{0}-J_{0}=(\Delta+M_{||})-(2R-M_{\perp}+r)=0~,
\end{equation}
where in the last step we make use of the Schur shortening conditions \eqref{schurshort}.  Hence the operators counted by \eqref{SSchurdef} are of $(c,c)$ type.  The universal variable $q$ appearing in the weights states according to the $2d$ charge  $C$ which arises from the commutant of the $(2,2)$ superalgebra inside the $4d$ $\mathcal{N}=2$ algebra in \eqref{conf2d}.

In the case of a conformal field theory with a conformally invariant surface defect we may use the state operator correspondence to related the index $\mathcal{I}_{\mathbb{S}}$ to a partition function on $S^{3}\times S^{1}$.  In this frame, the surface defect $\mathbb{S}$ wraps a great circle and extends along time.  Thus, the defect modifies the Hilbert space in radial quantization.  The local operators in the sum \eqref{SSchurdef} are states in this defect Hilbert space and the index is a supertrace.

For lagrangian theories it is straightforward to evaluate the defect index $\mathcal{I}_{\mathbb{S}}(q)$ using the tools of the previous sections.  Let us consider the class of defects for which the coupling to the $4d$ bulk arises through gauging a $2d$ flavor symmetry using the bulk gauge fields.  In this case we may proceed as follows.
\begin{itemize}
\item Compute the specialized elliptic genus $\mathcal{G}_{(c,c)}(u)$ for the purely two-dimensional degrees of freedom treating the four-dimensional bulk fields as fixed.  The resulting expression depends on $4d$ vector multiplet fields that appear as the two-dimensional flavor variables $u$.
\item Integrate over the variables $u$ by inserting the defect contribution $\mathcal{G}_{(c,c)}(u)$ into the Schur index integrand \eqref{operatorSchur}.
\end{itemize}
\paragraph{Example: $\mathbb{CP}^{1}$ Sigma Model Coupled to $\frak{su}(2)$ SYM}\mbox{}\\

Let us consider the case of $\frak{su}(2)$ SYM coupled to the $\mathbb{CP}^{1}$ sigma model by gauging the two-dimensional flavor symmetry.  The ordinary Schur index $\mathcal{I}(q)$ was given in \eqref{su2schur}, while the specialized genus $\mathcal{G}_{(c,c)}(u)$ of the $\mathbb{CP}^{1}$ sigma model was written in \eqref{CP1special}.  The defect index is therefore
\begin{eqnarray}\label{su2localization}
\mathcal{I}_{\mathbb{S}}(q) & = & \oint \frac{du}{2\pi i u} \frac{(1-u^{2})(1-u^{-2})}{2}(qu^{2};q)_{\infty}^{2}(q;q)_{\infty}^{2}(qu^{-2};q)_{\infty}^{2}\mathcal{G}_{(c,c)}(u) \nonumber\\
& = &  \oint \frac{du}{2\pi i u} \frac{(1-u^{2})(1-u^{-2})}{2}(qu^{2};q)_{\infty}^{2}(q;q)_{\infty}^{2}(qu^{-2};q)_{\infty}^{2}(-u^{2}-u^{-2}) \label{cp1surface}\\
& = &1+2\sum_{n=1}^{\infty}q^{n^{2}}=\theta_{3}(2\tau)~. \nonumber
\end{eqnarray}
Note, curiously, that like the Schur index \eqref{su2schur} the above has simple modular properties.  In section \ref{sec:simex} we will reproduce this formula for $\mathcal{I}_{\mathbb{S}}(q)$ using $2d$-$4d$ BPS particles.

\section{Infrared Formulas for $2d$-$4d$ Schur Indices}

In this section we conjecture infrared formulas for $2d$-$4d$ Schur indices.  Our expressions generalize those of \cite{Cordova:2015nma,Cordova:2016uwk} and express the index in terms of contributions from the BPS states on the Coulomb branch.  These formulas intertwine the Cecotti-Vafa formalism \cite{Cecotti:1992rm} for specialized elliptic genera with the results of \cite{Cordova:2015nma}.

\subsection{$2d$ BPS Solitons and Cecotti-Vafa Formulae}\label{sec:CV}

We begin with a review of the Cecotti-Vafa formalism which expresses the $2d$ specialized genus in terms of the BPS soliton spectrum after relevant deformation.

Two-dimensional $(2,2)$ theories admit a variety of supersymmetric relevant deformations.
\begin{itemize}
\item Deformations by superpotential ($W$) or twisted superpotential ($\widetilde{W}$) terms. 

\item Deformations by untwisted mass parameters ($m$) or twisted mass parameters ($\widetilde{m}$) \cite{Hanany:1997vm} which couple respectively to twisted current multiplets (charge $\widetilde{\gamma}$) or current multiplets (charge $\gamma$) .
\end{itemize}
In general these deformations break part of the $R$-symmetry of the conformal field theory.  The superpotential and mass break $R_{V},$ but preserve $R_{A}$, while the twisted superpotential and twisted mass break $R_{A},$ but preserve $R_{V}.$

We consider a theory in the presence of such deformations and assume that the infrared dynamics is gapped with a finite set of $N$ vacua labelled by an index $i$.  We are concerned with the spectrum of massive particles. These may reside in a single vacuum $i$ or they may be solitons interpolating between distinct vacua $i$ and $j$.  In the presence of the deformations the superalgebra \eqref{qcharges} in the particle sector $i-j$ is centrally extended as\footnote{We here use $\gamma$ and $\widetilde{\gamma}$ to denote multiple global charges, each of which may support an independent mass parameter.} 
\begin{equation}\label{massivealgebra}
\{Q_{+},Q_{-}\}=W_{i}-W_{j}+m \widetilde{\gamma}~, \hspace{.5in}\{Q_{+},\bar{Q}_{-}\}=\widetilde{W}_{i}-\widetilde{W}_{j}+\widetilde{m} \gamma~,
\end{equation}
where the subscripts indicate evaluation of the (twisted) superpotential in the specified vacuum.

Let us now specialize to a class of deformations compatible with the existence of a specialized genus.  Without loss of generality we consider the $(c,c)$ specialization which requires an unbroken vectorial $R$-symmetry $R_{V}$.  This implies that the superpotential deformation $W$ and untwisted mass $m$ vanish.  Then, there is a BPS bound in the $i-j$ particle sector 
\begin{equation}
M\geq |\mathcal{\widetilde{Z}}_{ij}|=|\widetilde{W}_{i}-\widetilde{W}_{j}+\widetilde{m} \gamma|~.
\end{equation}
Particles saturating the bound are in short representations of the superalgebra.  We count these short multiplets using an index \cite{Cecotti:1992qh}.  For BPS solitons interpolating between distinct vacua we define   
\begin{equation}
\mu_{ij,\gamma}=-\mu_{ji,-\gamma}= \mathrm{Tr}_{ij,\gamma}\left((-1)^{F_{2d}}F_{2d}\right)~.
\end{equation}
Here the trace is over the Hilbert space of one-particle states interpolating between the $i$-th and $j$-th vacuum carrying flavor charge $\gamma$.  The relationship between $\mu_{ij}$ and $\mu_{ji}$ follows from the fact that these sectors are related by CPT and hence carry opposite $F_{2d}$ charge.

Similarly we may count BPS particles in the vacuum $i$.
\begin{equation}
\omega_{i}(\gamma)=-\omega_{i}(-\gamma)= \mathrm{Tr}_{ii,\gamma}\left((-1)^{F_{2d}}F_{2d}\right)~.
\end{equation}
Unlike the solitons counted by $\mu_{ij},$ the particles counted by $\omega_{i}(\gamma)$ can only exist in the case where we have activated twisted mass parameters and they necessarily carry flavor charges.  The relationship between the index for charge $\gamma$ and the index for charge $-\gamma$ is again due to the fact that these sectors are related by CPT. 

In the absence of twisted masses it is common to find chambers where the given $2d$ massive theory has only a finite number of BPS solitons.  By contrast, when we activate twisted mass parameters there may be BPS particles and these tend to be accompanied by and infinite number of BPS solitons whose central charges accumulate at the particle rays.  This is qualitatively similar to the situation for higher spin massive BPS particles in $4d$ $\mathcal{N}=2$ theories.

\subsubsection{$2d$ Wall-Crossing Formula and Specialized Indices}

The $2d$ BPS spectra may jump across walls of marginal stability in the parameter space of relevant deformations.  The changes in the spectrum are governed by the Cecotti-Vafa wall-crossing formula \cite{Cecotti:1992rm}, which we review here.\footnote{These formulae have been given a categorified interpretation in \cite{Gaiotto:2015aoa}, which helps explain why UV local operators can be associated to certain sequences of IR solitons. }
  
The formula asserts that a certain $N\times N$ matrix (with again $N$ the number of vacua) is independent of the chamber.  The matrix is constructed as a product of factors, one for each BPS particle sector.  As parameters change the decomposition of the matrix will jump but the final answer will stay the same.  

Let $\delta^{i}_{j}$ denote the identity matrix and $e^{i}_{j}$ a matrix whose only non-vanishing is a one in the  $i$-th row and $j$-th column.  The elementary building block matrices in the wall-crossing formula are determined from the indices as follows
\begin{equation}
(ij)~\mathrm{solitons}\rightarrow S^{2d}_{ij;\gamma}\equiv\delta^{i}_{j}-\mu_{ij,\gamma}u(\gamma)e^{i}_{j}~, \hspace{.5in} (i,\gamma)~\mathrm{particles} \rightarrow K^{2d}_{\gamma}\equiv \sum_{i=1}^{N}(1-u(\gamma))^{-\omega_{i}(\gamma)}e^{i}_{i}~. \label{2dfactors}
\end{equation}
Here $u(\gamma)$ denotes the flavor fugacity for the global charge $\gamma$.  

We use these ingredients to define generating functions ($N\times N$ matrices) $\mathcal{S}^{2d}_{\vartheta_1,\vartheta_2},$ where $[\vartheta_1, \vartheta_2)$ is an angular sector in the plane of twisted central charges.
\begin{equation}
\mathcal{S}^{2d}_{\vartheta_1,\vartheta_2}(u)= :\prod_{\arg(\mathcal{\widetilde{Z}})\in [\vartheta_1 , \vartheta_2)}^{\curvearrowleft}
S^{2d}_{ij;\gamma} K^{2d}_{\gamma'}:
~.
\end{equation}
Here the BPS factor matrices take the form of \eqref{2dfactors} depending on whether they are solitons or particles, and the various factors in the product are ordered according to increasing phase of the twisted central charge $\mathcal{\widetilde{Z}}$ indicated by the normal ordering notation.  

With these ingredients, we can now formulate the wall-crossing formula.  Indeed, the statement is that for any angular sector the $N\times N$ matrix $\mathcal{S}^{2d}_{\vartheta_{1}, \vartheta_{2}}$ is independent of the relevant deformation parameters, provided no solitons exit the wedge $(\vartheta_{1}, \vartheta_{2})$.  This holds even though the BPS spectrum itself depends on the parameters.  In particular the matrix $\mathcal{S}^{2d}_{\vartheta, \vartheta+\pi}$ includes a contribution from all the sectors not related by CPT and hence may be viewed as a generating function for the spectrum.

The spectrum generator is chamber independent, but its particular matrix representation depends on a choice of basis in the set of $N$ vacua.  We can eliminate this feature, and extract quantities that dependent only on the ultraviolet theory but not on the chamber, by passing to appropriate functions of the eigenvalues.  A natural candidate is the trace of the product of particle and antiparticle generators.  The Cecotti-Vafa formula asserts that this is equal to the specialized elliptic genus  
\begin{equation}
\mathcal{G}_{(c,c)}(u)=\mathrm{Tr}\left[\mathcal{S}^{2d}_{\vartheta, \vartheta+\pi}\mathcal{S}^{2d}_{\vartheta+\pi, \vartheta+2\pi}\right]~. \label{cvindex}
\end{equation}
Note in particular that the product of BPS factors is now over the full circle in the twisted central charge space and that therefore the above does not depend on the initial angle $\vartheta$.  We will see examples of this in sections \ref{sec:airy} and \ref{sec:cp1}.

\subsubsection{Extension To $2d$ Framed BPS States}
\label{pure2dframed}

The $2d$ wall-crossing formalism may be enriched by studying the theory in the presence of BPS boundary conditions.   We review these $2d$ BPS boundary conditions as a preparation for the relation between surface and line defects in section \ref{sec:linesurrel}. Such boundary conditions are labelled by an angle $\vartheta$ that determines supercharges they preserve.  There are right boundary conditions $B^{\alpha}(\vartheta)$ and left boundary conditions $B_{\alpha}(\vartheta)$.

In the infrared the boundary conditions support framed BPS states where at $\pm \infty$ we place a given massive vacuum $i$.  We may count framed BPS states using indices in each sector
\begin{equation}
\chi^{\alpha, i}= \mathrm{Tr}_{i,B^{\alpha}(\vartheta)}\left((-1)^{F_{2d}}\right)~, \hspace{.5in}\chi_{\alpha, i}= \mathrm{Tr}_{B_{\alpha}(\vartheta),i}\left((-1)^{F_{2d}}\right)~.
\end{equation}
Note that there is no momentum zero mode for framed BPS states, and hence the indices are not weighted with additional fermion number insertions. We assemble these framed indices into generating functions, which we represent concretely as $N$ component vectors
\begin{equation}
F(B^{\alpha},\vartheta)=\left(\begin{array}{c}\chi^{\alpha,1 } \\ \chi^{\alpha,2 } \\ \vdots \\ \chi^{\alpha,N }\end{array}\right)~,\hspace{.5in}F(B_{\alpha},\vartheta)=\left(\chi_{\alpha,1 } ~ \chi_{\alpha,2 } ~ \cdots ~ \chi_{\alpha,N }\right)~.
\end{equation}

Almost by definition, the framed BPS degeneracies control the Witten index of the theory compactified on a segment with boundary conditions 
$B_\alpha$ and $B^\beta$: it must coincide with the inner product 
\begin{equation}
(B_\alpha, B^{\beta}) \equiv \chi_{\alpha}^\beta = F(B_{\alpha},\vartheta)F(B^{\beta},\vartheta)~.
\end{equation}
The results of \cite{Gaiotto:2015aoa} imply some simple generalizations of the Cecotti-Vafa formula which have a similar form: 
the index counting chiral operators living at the tip of a wedge of angular width $\vartheta' - \vartheta$, with boundary conditions  
$B_\alpha(\vartheta)$ and $B^\beta(\vartheta')$ is 
\begin{equation}
\mathcal{G}_{c}[B_\alpha(\vartheta),B^\beta(\vartheta')] = F(B_{\alpha},\vartheta)\mathcal{S}^{2d}_{\vartheta, \vartheta'} F(B^{\beta},\vartheta)~.
\end{equation}
In particular, local operators intertwining $B_\alpha$ and $B_\beta$ boundary conditions take the form 
\begin{equation}
\mathcal{G}_{c} [B_\alpha(\vartheta),B_\beta(\vartheta)]= F(B_{\alpha},\vartheta)\mathcal{S}^{2d}_{\vartheta, \vartheta+\pi} \overline{ F(B_{\beta},\vartheta)}~,
\end{equation}
where one uses CPT conjugation to map a left boundary condition $B_\alpha(\vartheta)$ to a right boundary condition at $\vartheta + \pi$. 

It is instructive to study the dependence of the framed BPS degeneracies as the angle $\vartheta$ defining the boundary condition is varied.  When $\vartheta$ crosses rays containing $2d$ BPS particles or solitons the framed spectrum may change according to 
\begin{equation}
\mathcal{S}^{2d}_{\vartheta, \vartheta'}F(B^{\alpha},\vartheta')=F(B^{\alpha},\vartheta)~, \hspace{.5in}F(B_{\alpha},\vartheta)\mathcal{S}^{2d}_{\vartheta, \vartheta'}=F(B_{\alpha},\vartheta')~.\label{2dframedwc}
\end{equation}
Conversely, if we know how a $\vartheta \to \vartheta'$ rotation acts on a sufficiently large set of UV boundary conditions 
and we know how to compute framed BPS degeneracies for UV boundary conditions, we can deduce the matrices $\mathcal{S}^{2d}_{\vartheta, \vartheta'}$  and the BPS spectrum of the theory. 

A consequence of these constructions is that there is monodromy in set of boundary conditions as $\vartheta\rightarrow \vartheta+2\pi$.
We denote this action abstractly by an operator $\mathcal{R}_{2\pi}$
\begin{equation}
B_{\alpha}(\vartheta+2\pi)=\left(\mathcal{R}_{2\pi}\circ B_{\alpha}\right)(\vartheta)~, \hspace{.5in}B^{\alpha}(\vartheta+2\pi)=\left(\mathcal{R}_{2\pi}\circ B^{\alpha}\right)(\vartheta)~.
\end{equation}

There is a nice relation between the specialized elliptic genus, the $\mathcal{R}_{2\pi}$ operation and the Witten indices on segments. 
Indeed, we can write 
\begin{equation}
(\mathcal{R}_{2\pi} \circ B_\alpha,  B^{\beta}) = F(B_{\alpha},\vartheta)\mathcal{S}^{2d}_{\vartheta, \vartheta + 2 \pi} F(B^{\beta},\vartheta)
\end{equation}
Given any $N$ boundary conditions whose framed BPS degeneracies are linearly independent, we can compute the trace in the corresponding basis: 
\begin{equation} \label{eq:unwrapped}
\mathcal{G}_{(c,c)}(u)=\sum_{\alpha, \beta} (\chi^{-1})^{\alpha}_\beta(\mathcal{R}_{2\pi} \circ B_\alpha,  B^{\beta}) ~.
\end{equation}

This expression is a toy model for the ``unwrapping'' process promised in the introduction. 
Given a ``good enough'' basis of left and right boundary conditions, generating the appropriate categories of branes for the theory,
we can express the identity interface as a deformation of a direct sum of products of left and right boundary conditions.\footnote{Theories with isolated massive vacua usually admit particularly nice bases of dual boundary conditions (``thimbles''), 
for which $\chi_{\alpha}^\beta =\delta_\alpha^\beta$. }
Whenever we do that, the framed BPS degeneracies for the identity interface are decomposed accordingly: 
\begin{equation}
\mathrm{Id}_{N\times N} = \sum_{\alpha, \beta} (\chi^{-1})^{\alpha}_\beta F(B^{\beta},\vartheta) \otimes F(B_{\alpha},\vartheta)~,
\end{equation}
where $\mathrm{Id}_{N\times N}$ is the $N\times N $ identity matrix.

The bulk local operators can be thought of as the operators which live at the end of the identity interface:
\begin{equation}
\mathcal{G}_{(c,c)}(u)=\mathrm{Tr}\left[\mathcal{S}^{2d}_{\vartheta, \vartheta+\pi}\mathcal{S}^{2d}_{\vartheta+\pi, \vartheta+2\pi} \mathbf{1} \right]~,
\end{equation}
and the index decomposes as a sum of indices for wedges of width $2 \pi$ with boundary conditions $B^{\beta}$ and $B_{\alpha}$.
The identity with \ref{eq:unwrapped} corresponds to a continuous interpolation between a wedge of width $2\pi$ and a wedge of width $0$.

\subsubsection{The Airy Example}
\label{sec:airy}

As an example let us consider the simplest twisted Landau-Ginzburg theory.  In the ultraviolet the theory is defined by a homogeneous cubic twisted superpotential.  We may activate a relevant twisted superpotential parameter and flow to infrared
\begin{equation}
\widetilde{W}_{UV}=\widetilde{\Phi}^{3}\longrightarrow \widetilde{W}_{IR}=\widetilde{\Phi}^{3}-\Lambda \widetilde{\Phi}~.
\end{equation}
In the infrared the theory is gapped and has two vacua.  There is a single BPS soliton hence
\begin{equation}
\mu_{12}=-\mu_{21}=1~.
\end{equation}
We can use this data to evaluate the specialized genus using the Cecotti-Vafa formula \eqref{cvindex}: 
\begin{equation}
\mathcal{G}_{(c,c)} = ~\text{Tr} \left[ \begin{pmatrix}1 & -1 \cr 0 & 1\end{pmatrix} \begin{pmatrix}1 & 0 \cr 1 & 1\end{pmatrix}\right] =1~,
\end{equation}
which reproduces the correct answer \eqref{mmspecialized} for the specialized  elliptic genus. 

Let us also discuss the boundary conditions and framed BPS states. Boundary conditions can be visualized as Lagrangian branes which 
go to infinity along three specific directions in the $\widetilde{\Phi}$ plane where the superpotential goes to $\infty$ in the $e^{i \vartheta}$ direction
(see \cite{Gaiotto:2015zna,Gaiotto:2015aoa} and references therein for details). 

There are three naturally defined right boundary conditions $B_1$, $B_2$ and $B_3$. In an appropriate chamber, their framed BPS spectra are \begin{equation}
F(B_1) = (1 \quad 0) ~, \hspace{.5in}F(B_2) = (0 \quad -1)~,\hspace{.5in}F(B_3) = (-1 \quad -1)~.
\end{equation}
Using the matrix $\mathcal{S}_{\vartheta, \vartheta+2\pi}$ appearing in the trace above, one may verify that the action of the monodromy $\mathcal{R}_{2\pi}$ is 
\begin{equation}
\mathcal{R}_{2\pi} \circ B_{1}=B_{2}~, \hspace{.5in}\mathcal{R}_{2\pi} \circ B_{2}=B_{3}~, \hspace{.5in}\mathcal{R}_{2\pi} \circ B_{3}=\mathbb{C}^{[1]}( B_{1})~,
\end{equation}
where in the last line the notation $\mathbb{C}^{[n]}(B_{\alpha})$ indicates the boundary condition $B_{\alpha}$ but with a shift in fermion number by $n$ units (in this case accounted for by a sign in the framed degeneracies). This has a simple geometric interpretation: as $\vartheta$ varies, the 
asymptotic directions in the $\widetilde{\Phi}$ plane rotate accordingly. 

Similarly, there are left boundary conditions $B^1$, $B^2$, $B^3$ with framed BPS degeneracies
\begin{equation}
F(B^{1})=\left(\begin{array}{c} 1 \\ 0 \end{array}\right)~, \hspace{.5in}F(B^{2})=\left(\begin{array}{c} 0 \\ -1 \end{array}\right)~, \hspace{.5in}F(B^{2})=\left(\begin{array}{c} 1 \\ -1 \end{array}\right)~.
\end{equation}
These are also cyclically permuted by the monodromy matrix $\mathcal{R}_{2\pi}$ up to appropriate fermion number shifts.

We may pick two of these right boundary conditions, together with dual right boundary conditions to construct a resolution of the identity interface.  For instance a simple choice is to take $B^{1}$ and $B^{2}$ which has as a dual basis $B_{1}$ and $B_{2}$.  We can then compute
\begin{equation}
F(B^{1})F(B_{1})+F(B^{2})F(B_{2})=\left(\begin{array}{cc} 1 & 0\\ 0 & 1\end{array}\right)~.
\end{equation}

\subsubsection{The $\mathbb{CP}^1$ Sigma Model}
\label{sec:cp1}

A richer example is the $\mathbb{CP}^1$ sigma model, whose BPS spectrum has been studied in \cite{Cecotti:1992rm,Hanany:1997vm,Dorey:1998yh,Gaiotto:2011tf}.  This theory is asymptotically free and has two massive vacua in the infrared.  In the absence of twisted masses, the spectrum consists of two BPS solitons interpolating between the two vacua.  These solitons transform as a doublet under the $\frak{su}(2)$ flavor symmetry.  The BPS indices are thus
\begin{equation}
\mu_{12}=-\mu_{21}=u+u^{-1}~,
\end{equation} 
where $u$ is the $\frak{su}(2)$ flavor fugacity. 
We can use this data to compute the specialized genus using the Cecotti-Vafa formula \eqref{cvindex}
\begin{equation}
\mathcal{G}_{(c,c)}(u) = ~\text{Tr} \left[ \begin{pmatrix}1 & -u - u^{-1} \cr 0 & 1\end{pmatrix} \begin{pmatrix}1 & 0 \cr u + u^{-1} & 1\end{pmatrix}\right] = - u^2 - u^{-2} = 1 - \chi_3(u)~.
\end{equation}
This reproduces the explicit result \eqref{CP1special} obtained by localization using the ultraviolet lagrangian.

Next, we can add twisted masses. If the masses are small, the only effect is to split the two BPS rays for the solitons of different flavor charge.  The associated wall-crossing identity is simply
\begin{equation}
\mathcal{S}_{\vartheta, \vartheta+\pi}=\begin{pmatrix}1 & -u - u^{-1} \cr 0 & 1\end{pmatrix}= \begin{pmatrix}1 & -u  \cr 0 & 1\end{pmatrix} \begin{pmatrix}1 & -u^{-1} \cr 0 & 1\end{pmatrix}~.
\end{equation}
As we increase the twisted mass, rays from BPS solitons of opposite topological charge but equal flavor charge meet in the twisted central charge plane, and wall-crossing occurs.  The associated matrix identity is
\begin{equation}
\begin{pmatrix}1 & -u^{-1} \cr 0 & 1\end{pmatrix} \begin{pmatrix}1 & 0 \cr u^{-1}  & 1\end{pmatrix} = \prod_{n\geq0} \begin{pmatrix}1 & 0 \cr u^{-2n-1}  & 1\end{pmatrix} \begin{pmatrix}1-u^{-2} & 0 \cr 0 & (1-u^{-2})^{-1}\end{pmatrix} \prod_{n\geq0}\begin{pmatrix}1 & -u^{-2n-1} \cr 0 & 1\end{pmatrix}~.
\end{equation}
In this ``weak coupling'' chamber there is an infinite tower of solitons of charges $2n+1$ and a single BPS particle of charge $2$ in each vacuum. 

This model has a variety of interesting boundary conditions (corresponding to B-branes in the sigma model) \cite{Gaiotto:2015aoa}. Starting from boundary conditions $B_0$ and $B_1$ with $F(B_1) = (1 \quad 0)$ and $F(B_0) = (0 \quad 1)$  we can obtain an infinite family of boundary conditions $B_n$ by acting with the monodromy $\mathcal{R}_{2 \pi}:$  
\begin{equation}
\mathcal{R}_{2\pi}^{n}\circ B_{0}\equiv B_{2n}~,\hspace{.5in}\mathcal{R}_{2\pi}^{n}\circ B_{1}\equiv B_{2n+1}~.
\end{equation}
These boundary conditions preserve the $\frak{su}(2)$ global symmetry and have framed degeneracies
\begin{equation}
F(B_k) = \begin{cases}(-1)^{\lfloor\frac{k-1}{2}\rfloor}(\chi_{k}(u) \quad \chi_{k-1}(u)) &k>0~, \\
(-1)^{\lfloor\frac{k}{2}\rfloor}(\chi_{-k}(u) \quad \chi_{-k+1}(u)) &k\leq0~,  \end{cases}
\end{equation}
where $\chi_{n}(u)$ is the character of the $n$-dimensional representation.  Geometrically these boundary conditions are associated to line bundles $\mathcal{O}(n)$ on $\mathbb{CP}^1$. There are similar families of dual branes $B^{n}$.

There are also boundary conditions $B_\pm$ localized at either pole of $\mathbb{CP}^1$ which break the $\frak{su}(2)$ global symmetry to a Cartan $\frak{u}(1)$. These are eigenbranes under ${\mathcal R}_{2 \pi}$. They have $F(B_\pm) = ( 1 \quad  u^{\pm 1})$ and ${\mathcal R}_{2 \pi}$ eigenvalues $-u^{\mp 2}$. Using $B_\pm$ as a basis for a resolution of the identity yields a simple calculation of the specialized genus.

\subsection{$4d$ BPS Particles and Infrared Formulas for Schur Indices}\label{sec:CS}

In this section we review the results of \cite{Cordova:2015nma} that reconstruct the Schur index of $4d$ $\mathcal{N}=2$ theories from their Coulomb branch data and BPS particles.  These formulas are conceptually quite similar to the $2d$ Cecotti-Vafa formalism described in the previous sections.  However, the massless degrees of freedom in the infrared together with the additional quantum numbers of four-dimensional physics together yield a promotion of the wall-crossing data from $N\times N$ matrices in the two-dimensional case to infinite-dimensional matrices in the four-dimensional case. 

We begin with a general four-dimensional $\mathcal{N}=2$ theory and activate Coulomb branch parameters to generate an RG flow.\footnote{Here we include the possibility of activating flavor masses which appear as scalars in  non-dynamical vector multiplets.}  In the infrared the physics is that of an abelian gauge theory with gauge group $\frak{u}(1)^{r}$ where $r$ is the rank of the theory.  There is an integral lattice $\Gamma$ of electromagnetic and flavor charges.  The lattice has several structures:
\begin{itemize}
\item A bilinear, antisymmetric, integer valued Dirac pairing $\langle \cdot, \cdot \rangle.$  The flavor charges define a sublattice $\Gamma_{f}\subset \Gamma$ which have trivial pairings with all other charges.  The pairing is non-degenerate on the quotient $\Gamma/\Gamma_{f}$.
\item A linear central charge function $\mathcal{Z}:\Gamma\rightarrow \mathbb{C},$ determined by the Seiberg-Witten formalism.
\end{itemize}

For each charge $\gamma\in \Gamma$ there can be BPS particles saturating the bound $\mathrm{mass}\geq |\mathcal{Z}(\gamma)|$.  These may be counted using an appropriate index.  In general the BPS particles are a representation of the super little group $\frak{su}(2)_{J}\times \frak{su}(2)_{R}$, where $J$ denotes the spin.  This representation takes the form
\begin{equation}
\left[(\mathbf{2},\mathbf{1})\oplus (\mathbf{1}, \mathbf{2})\right]\otimes h_{\gamma}~,
\end{equation}
where in the above the factor in brackets may be viewed as the center of mass and the vector space $h_{\gamma},$ the internal degrees of freedom.  We require the degeneracies $\Omega_{n}(\gamma)$ defined as\footnote{The indices $\Omega_{n}(\gamma)$ are in fact all positive as a consequence of the absence of exotics, i.e. BPS particles carrying non-vanishing $\frak{su}(2)_{R}$ charge \cite{CD}.}
\begin{equation}
\mathrm{Tr}_{h_{\gamma}}\left[y^{J}(-y)^{R}\right]=\sum_{n\in \mathbb{Z}}\Omega_{n}(\gamma)y^{n}~.
\end{equation}
Ordinary hypermultiplets thus contribute only to $\Omega_{0}(\gamma)$ while non-vanishing $\Omega_{n}(\gamma)$ with $|n|>0$ imply the existence of higher-spin BPS particles.  Such higher-spin BPS particles are typically accompanied by infinite cohorts of hypermultiplets.

\subsubsection{Wall-Crossing Formulas and the IR Schur index}

As Coulomb branch parameters are varied, the BPS spectrum may jump.  These discontinuities in the indices $\Omega_{n}(\gamma)$ are encoded by the Kontsevich-Soibelman wall-crossing formula.  

We construct wall-crossing invariant quantum generating functions of the BPS particles $\mathcal{S}^{4d}_{\vartheta_1, \vartheta_2}(q)$.  The generating function is presented as a product of factors which arise from BPS particles with central charge $\mathcal{Z}(\gamma)$ in the angular sector $[\vartheta_1 , \vartheta_2).$  In general $\mathcal{S}^{4d}_{\vartheta_1, \vartheta_2}(q)$ is only defined for  $\vartheta_2- \vartheta_1\leq \pi$.  The statement of the wall-crossing formula is that $\mathcal{S}^{4d}_{\vartheta_1, \vartheta_2}(q)$ is chamber independent, even though its presentation as a product is chamber dependent \cite{Kontsevich:2008fj}.

To concretely define the generating function $\mathcal{S}^{4d}_{\vartheta_1, \vartheta_2}(q),$ we must introduce a quantum torus algebra associated to the charge lattice $\Gamma$.  For each charge vector $\gamma\in \Gamma$ we introduce a variable $X_{\gamma}.$  They may be multiplied as
\begin{equation}
X_\gamma X_{\gamma'} = q^{\frac12\langle \gamma, \gamma' \rangle} X_{\gamma + \gamma'}~. \label{torus}
\end{equation}
Physically these variables define the algebra of line defects in the infrared $\frak{u}(1)^{r}$ gauge theory.  (See e.g \cite{Gaiotto:2010be} for an extended discussion).

The generating function $\mathcal{S}^{4d}_{\vartheta_1, \vartheta_2}(q)$ is a semi-infinite power series in quantum torus variables $X_{\gamma}$ 
\begin{equation}
\mathcal{S}^{4d}_{\vartheta_1, \vartheta_2}(q) \equiv \sum_{\gamma | \arg(\mathcal{Z}_{\gamma})\in [\vartheta_1 , \vartheta_2)} s^\gamma_{\vartheta_1, \vartheta_2}(q) X_\gamma~,
\end{equation}
where $s^\gamma_{\vartheta_1, \vartheta_2}(q)$ is a Laurent series in $q$.  By definition they satisfy the composition rule
\begin{equation}
\mathcal{S}^{4d}_{\vartheta_1, \vartheta_2}(q) \mathcal{S}^{4d}_{\vartheta_2, \vartheta_3}(q) = \mathcal{S}^{4d}_{\vartheta_1, \vartheta_3}(q)~, 
\end{equation}
where the multiplication above is defined using the quantum torus algebra \eqref{torus}.  The $\mathcal{S}^{4d}_{\vartheta_1, \vartheta_2}(q)$ are constructed out of individual BPS particle rays as
\begin{equation}
\mathcal{S}^{4d}_{\vartheta_1,\vartheta_2}(q)= \prod_{\gamma| \arg(\mathcal{Z}_{\gamma})\in [\vartheta_1 , \vartheta_2)}^{\curvearrowleft} K^{4d}(q;X_\gamma;\Omega_{j}(\gamma))~,
\end{equation}
with 
\begin{equation}
K^{4d}_\gamma(q;\Omega_{j}(\gamma))=\prod_{n\in \mathbb{Z}}E_{q}((-1)^{n}q^{n/2}X_{\gamma})^{(-1)^{n}\Omega_{n}(\gamma)}~.
\end{equation}
We will often omit the dependence of $K^{4d}_\gamma$ on $q$ and $\Omega_j$.  
Here $E_q(z)$ is the quantum dilogarithm defined as
\begin{equation}
E_{q}(z) = (-q^{\frac12} z;q)^{-1}_\infty = \prod_{i=0}^{\infty}(1+q^{i+\frac{1}{2}}z)^{-1}=\sum_{n=0}^{\infty}\frac{(-q^{\frac{1}{2}}z)^{n}}{(q)_{n}}~. \label{eqdef}
\end{equation}
The statement of the wall-crossing formula is then simply that $\mathcal{S}^{4d}_{\vartheta_1, \vartheta_2}(q)$ is independent of moduli.  

With these ingredients we can formulate the conjecture of \cite{Cordova:2015nma} for the Schur index $\mathcal{I}(q)$ in terms of the infrared Coulomb branch data as
\begin{equation}\label{irshur}
\mathcal{I}(q) = (q)_\infty^{2r}~\text{Tr} \left[ \mathcal{S}^{4d}_{\vartheta,\vartheta+\pi}(q) \mathcal{S}^{4d}_{\vartheta+\pi,\vartheta+2\pi}(q)\right]~.
\end{equation}
Here the trace operation appearing above should be thought of as an inner product between two well-defined elements in vector spaces of semi-infinite sums:
\begin{equation}
\mathcal{I}(q) = (q)_\infty^{2r}~\sum_{\gamma} s^\gamma_{\vartheta,\vartheta+\pi}(q)s^{-\gamma}_{\vartheta+\pi,\vartheta+2\pi}(q)~. \label{tracedef}
\end{equation} 
The independence of $\mathcal{I}(q)$ from $\vartheta$ is not obvious, and appears to be a non-trivial constraint on the BPS spectrum of well-defined ${\cal N}=2$ theories.  

More generally we can also obtain the dependence on flavor fugacities $x$ by identifying  the commuting elements $X_{\gamma_f}$ of the quantum torus algebra. In this case the sum in \eqref{tracedef} is only over the nonvanishing electromagnetic charges.

The physical interpretation of this IR formula was reviewed in section \ref{sec:IRintro}.

\subsubsection{Extension to Line Defects and Framed BPS States}
\label{sec:linedef}

The framework of the previous section carries over if the theory is probed by BPS line defects $L_{i}$.    These line defects carry a choice of phase $\vartheta$ which determines the unbroken supersymmetry algebra.

When these defects extend along time, they modify the Hilbert space.  In the infrared this leads to a new class of BPS states, framed BPS states, which may be viewed physically as particles bound to the line defect.  We encode these defects in an index
\begin{equation}
\fro(L,\vartheta,\gamma,q)=\mathrm{Tr}_{h_{L,\gamma}}\left(q^{J}(-q)^{R}\right)~,
\end{equation}
and package the entire collection of indices into a  framed BPS characters, $F(L,\vartheta)$, which is a finite Laurent polynomial in the $X_\gamma$ \cite{Gaiotto:2010be}
\begin{equation}
 F(L,\vartheta)=\sum_{\gamma\in \Gamma}\fro(L,\vartheta,\gamma,q) X_{\gamma}~.
\end{equation} 

Using this data, we can compute the Schur index with $n+m$ line defects insertions $L_i^{\vartheta_i}$ ordered as 
\begin{equation}
\vartheta\leq \vartheta_1< \cdots<\vartheta_{n} \leq \vartheta+\pi \leq \vartheta_{n+1}< \cdots<\vartheta_{n+m}\leq \vartheta+2 \pi 
\end{equation}
via the formula \cite{Cordova:2016uwk}
\begin{equation}\label{irshurl}
\mathcal{I}_{(L_i^{\vartheta_i})}(q) = (q)_\infty^{2r}~\text{Tr} \left[  \mathcal{S}^{(L_i^{\vartheta_i})}_{\vartheta,\vartheta+\pi}(q) \mathcal{S}^{(L_i^{\vartheta_i})}_{\vartheta+\pi,\vartheta+2\pi}(q) \right]~,
\end{equation}
where in the above
\begin{align}
\mathcal{S}^{(L_i^{\vartheta_i})}_{\vartheta,\vartheta+\pi}(q) &= \mathcal{S}^{4d}_{\vartheta,\vartheta_1}(q) \left[\prod_{i< n} F[L_i,\vartheta_i]  \mathcal{S}^{4d}_{\vartheta_i,\vartheta_{i+1}}(q)\right] F[L_n,\vartheta_n]   \mathcal{S}^{4d}_{\vartheta_n,\vartheta+\pi}(q)~,  \cr\mathcal{S}^{(L_i^{\vartheta_i})}_{\vartheta+\pi,\vartheta+2\pi}(q)  &= \mathcal{S}^{4d}_{\vartheta+\pi,\vartheta_{n+1}}(q)  \left[\prod_{n<i<m} F[L_i,\vartheta_i]  \mathcal{S}^{4d}_{\vartheta_i,\vartheta_{i+1}}(q)\right] F[L_m,\vartheta_m] \mathcal{S}^{4d}_{\vartheta_m,\vartheta+2\pi}(q) ~.
\end{align}
The phase attached to a line defect can be continuously deformed, without changing the Schur index. This is due to the framed wall-crossing formula \cite{Gaiotto:2010be} which governs the jumps in the framed indices $\fro(L,\vartheta,\gamma,q)$ as the angle $\vartheta$ crosses BPS rays
\begin{equation}
F(L,\vartheta)\mathcal{S}^{4d}_{\vartheta, \vartheta'}(q) = \mathcal{S}^{4d}_{\vartheta, \vartheta'}(q)F(L,\vartheta')~. \label{4dframedwc}
\end{equation}

Motivated by relations between lines and surfaces described in section \ref{sec:linesurrel}, let us also describe the non-trivial monodromies that are possible as the phase of the line defect is shifted by $2\pi$. The map $L^\vartheta \to L^{\vartheta + 2 \pi}$ reflects the presence or lack of $\frak{u}(1)_{r}$ symmetry in the theory:
\begin{itemize}
\item For asymptotically free theories, the $\frak{u}(1)_r$ rotation acts on the parameters of the theory. The map $L^\vartheta \to L^{\vartheta + 2 \pi}$
encodes a generalization of the Witten effect \cite{Witten:1979ey}. 
\item For  SCFTs with Coulomb branch operators of integral $\frak{u}(1)_r$ charges, the map $L^\vartheta \to L^{\vartheta + 2 \pi}$ is trivial. 
\item For  SCFTs with Coulomb branch operators of fractional $\frak{u}(1)_r$ charges (e.g. the Argyres-Douglas theories) with denominators dividing 
some common multiple $N$, the map $L^\vartheta \to L^{\vartheta + 2 \pi N}$ is trivial but $L^\vartheta \to L^{\vartheta + 2 \pi}$ is typically non-trivial. 
\end{itemize}
We again denote the result of transport from $\vartheta \to \vartheta + 2 \pi$ as the action of a map $\mathcal{R}_{2\pi}$:\footnote{It is tempting to formulate the monodromy operator $\mathcal{R}_{2\pi}$ using the full product $\mathcal{S}^{4d}_{\vartheta,\vartheta+\pi}(q) \mathcal{S}^{4d}_{\vartheta+\pi,\vartheta+2\pi}(q).$  However, in general such a formal product does not exist.  Instead, one can define $\mathcal{R}_{2\pi}$ as a composition of two parallel transports by angle $\pi$ using \eqref{4dframedwc}. }
\begin{equation}
L^{\vartheta + 2 \pi} = \left(\mathcal{R}_{2\pi}\circ L\right)^\vartheta~.
\end{equation}

\subsection{Refined $2d$-$4d$ Wall-Crossing Formula}\label{sec:2d4dWCF}

In this section we state the refined $2d$-$4d$ wall-crossing formula, generalizing the results of \cite{Gaiotto:2011tf} to include the grading of the $2d$-$4d$ particle spectrum by the universal flavor charge $C$.  The resulting formula is a natural extension of  the refined wall-crossing formula for  ordinary and framed BPS states \cite{Kontsevich:2008fj,Gaiotto:2010be,Dimofte:2009bv,Dimofte:2009tm}.

We begin with a $4d$ $\mathcal{N}=2$ theory coupled to a surface defect $\mathbb{S}$.  We now activate Coulomb branch parameters and flow to the infrared.  We assume that the defect theory is gapped with $N$ vacua.  The Coulomb branch parameters $u$ of the $4d$ vector multiplet appear as a twisted superpotential in the $2d$ defect theory, while the $4d$ electromagnetic charges $\Gamma$ appear in $2d$ as flavor charges.  

Let us now describe the various BPS objects that can appear in this coupled system.  There are of course bulk $4d$ BPS particles counted by indices $\Omega_n(\gamma)$ that are unaffected by the existence of the defect.  In addition, the defect supports BPS particles and solitons, but now such states may carry bulk charges $\Gamma$ and hence are referred to as $2d$-$4d$ BPS states.  Moreover, the $2d$-$4d$ BPS states may carry $\frak{u}(1)_{C}$ charge which appears in \eqref{conf2d} as the commutant of the $(2,2)$ superalgebra inside the $4d$ $\mathcal{N}=2$ algebra.  In the bulk
\begin{equation}
C=R-M_{\perp}~,
\end{equation}
thus counting $2d$-$4d$ BPS states refined by their $C$ charge is refining by $4d$ spin.  Note also that although it is abelian, the $C$ charge is quantized in half-integral units. We package these refined  $2d$-$4d$ BPS states into indices $\mu_{ij}(\gamma,n)$ and $\omega_{i}(\gamma,n)$ defined as
\begin{equation}
\mathrm{Tr}_{h_{ij,\gamma}}((-1)^{F_{2d}}F_{2d}\,y^{C})=\sum_{n\in \mathbb{Z}}\mu_{ij}(\gamma,n)y^{n/2}~, \hspace{.5in}\mathrm{Tr}_{h_{ii,\gamma}}((-1)^{F_{2d}}F_{2d}\, y^{C})=\sum_{n\in \mathbb{Z}}\omega_{i}(\gamma,n)y^{n/2}~.
\end{equation}
Notice that there is a certain ambiguity in attributing bulk gauge charges to $2d$ solitons. Formally speaking, the charges of $2d$ solitons are a ``torsor'' for the $4d$ charge lattice. 
Concretely, that means that there is a ``gauge freedom'' in defining charges for $2d$ solitons, involving shifts $\gamma \to \gamma + \gamma^{(i)} - \gamma^{(j)}$ 
of the charge of solitons between vacua $i$ and $j$. These shifts must be accompanied by a re-definition of the BPS degeneracies which is outlined below. 

The refined indices $\mu_{ij}(\gamma,n)$ and $\omega_{i}(\gamma,n)$ may jump as the Coulomb branch parameters are varied.  These jumps are governed by a wall-crossing formula that asserts that a certain operator constructed out of the spectrum is constant.  In this case the operator is a hybrid of the $2d$ operators appearing in section \ref{sec:CV} and the $4d$ operators appearing in section \ref{sec:CS} and is an $N\times N$ matrix whose entries are power series in the quantum torus variables $X_{\gamma}$.

We construct the wall-crossing operator with the following factor matrices.  The $2d$ solitons with topological charge $ij$ and $4d$ gauge charge $\gamma$ contribute factors of
\begin{equation}
S_{ij;\gamma}(q; \mu_{ij}(\gamma,k)) \equiv \delta^i_j -\sum_{k\in \mathbb{Z}} \mu_{ij}(\gamma,k) (-1)^{k}q^{\frac{k}{2}} X_{\gamma} e_{j}^{i}~. \label{2d4dS}
\end{equation}
Meanwhile, the $2d$ and $4d$ particles carrying charge $\gamma$ appear at the same phase in the central charge plane and contribute together as $K_\gamma \equiv K_\gamma^{4d}(q;\Omega(\gamma,n)) K^{2d}_\gamma(q;\omega_{i}(\gamma,n))$ where
\begin{equation}
K^{2d}_\gamma(q;\omega_{i}(\gamma,n)) \equiv \sum_i \prod_{n\in \mathbb{Z}} (1- (-1)^{n}q^{\frac{n}{2}}X_{\gamma})^{-\omega_i(\gamma,n)} e_{i}^{i}~.\label{2d4dK}
\end{equation}
We will often omit the dependence of $K_\gamma^{2d}$ and $S_{ij;\gamma}$ on $q$, $\mu_{ij}$, and $\omega_i$. 
The wall-crossing operator $\mathcal{S}_{\vartheta_1,\vartheta_2}(q)$ is then the phase ordered product
\begin{equation}\label{refinedwall2d}
\mathcal{S}^{2d-4d}_{\vartheta_1,\vartheta_2}(q)= :\prod_{ij,\gamma| \arg(\mathcal{Z})\in [\vartheta_1 , \vartheta_2)}^{\curvearrowleft} 
S_{ij;\gamma}K_{\gamma'}^{2d} K_{\gamma'}^{4d} :
~.
\end{equation}
The statement of the wall-crossing formula is that the operator $\mathcal{S}^{2d-4d}_{\vartheta_1,\vartheta_2}(q)$ is independent of Coulomb parameters as long as no BPS particle exits the wedge $(\vartheta_1,\vartheta_2)$.  

The signs accompanying powers of $q^{1/2}$ in the factor matrices of \eqref{2d4dS} and \eqref{2d4dK} require explanation.  These arise due to a mismatch between the $4d$ fermion number \eqref{4dfermion} and the $2d$ fermion number \eqref{2dfermion}.  Indeed, using the relationship of the $4d$ and $2d$ charges \eqref{cartanrels} and \eqref{cartanrels1}, as well as the Schur shortening conditions \eqref{schurshort} we can show that 
\begin{equation}
F_{2d}=F_{4d}+2C  \pmod 2~.
\end{equation}
Therefore objects carrying $C$ charge are counted with extra signs.  Note that we have used $M_{||}+M_\perp = -2j_1\in \mathbb{Z}$.

The ``gauge transformations'' \cite{Gaiotto:2011tf} which re-define the charges of 2$d$ solitons must be compatible with the wall-crossing formula. 
We claim that they are because they are equivalent to conjugating the wall-crossing operators by the diagonal matrix $\Delta$ with entries $X_{\gamma^{(i)}}$. 
Indeed, this conjugation acts appropriately on the 2$d$ gauge charges, and produces a shift in the BPS degeneracies, which refines 
the known unrefined formulae: 
\begin{equation}
\Delta S_{ij;\gamma}(q; \mu_{ij}(\gamma,k)) \Delta^{-1} \equiv \delta^i_j -\sum_{k\in \mathbb{Z}} \mu_{ij}(\gamma,k) (-1)^{k}q^{\frac{k}{2}} X_{\gamma^{(i)}} X_{\gamma} X_{-\gamma^{(j)}} e_{j}^{i}~,
\end{equation}
gives new 2$d$ soliton degeneracies determined by
\begin{equation}
 \sum_{k\in \mathbb{Z}} \mu'_{ij}(\gamma,k) (-1)^{k}q^{\frac{k}{2}} X_{\gamma+\gamma^{(i)}-\gamma^{(j)}} = \sum_{k\in \mathbb{Z}} \mu_{ij}(\gamma,k) (-1)^{k}q^{\frac{k}{2}} X_{\gamma^{(i)}} X_{\gamma} X_{-\gamma^{(j)}}~.
\end{equation}
Concretely,
\begin{equation}
 \mu'_{ij}(\gamma,k+  \langle \gamma^{(i)}+\gamma^{(j)},\gamma\rangle-   \langle \gamma^{(i)},\gamma^{(j)}\rangle) (-1)^{\langle \gamma^{(i)}+\gamma^{(j)},\gamma\rangle-   \langle \gamma^{(i)},\gamma^{(j)}\rangle}=\mu_{ij}(\gamma,k)   ~.
\end{equation}

Similarly the conjugation acts on the $K$ factors as
\begin{equation}
K^{4d}_\gamma(q;\Omega_{j}(\gamma)) K^{2d}_\gamma(q;\omega'_{i}(\gamma,n))=\Delta  K^{4d}_\gamma(q;\Omega_{j}(\gamma)) K^{2d}_\gamma(q;\omega_{i}(\gamma,n)) \Delta^{-1} ~.
\end{equation}
This gives new $\omega'_{i}(\gamma,n)$ in terms of the old $\omega_{i}(\gamma,n)$ and $\Omega_{j}(\gamma)$:
\begin{align}
\prod_{n\in \mathbb{Z}} &(1- (-1)^{n}q^{\frac{n}{2}}X_{\gamma})^{-\omega'_i(\gamma,n)- (-1)^n \sum_{i\geq 0} \Omega_{n-1-2i}(\gamma)} = \cr
&= \prod_{n\in \mathbb{Z}} (1- (-1)^{n}q^{\frac{n}{2}+ \langle \gamma^{(i)},\gamma\rangle}X_{\gamma})^{-\omega_i(\gamma,n)- (-1)^n \sum_{i\geq 0} \Omega_{n-1-2i}(\gamma)} ~,
\end{align}
and thus 
\begin{equation}
\omega'_i(\gamma,n+ 2 \langle \gamma^{(i)},\gamma\rangle)=\omega_i(\gamma,n)+ (-1)^n \left(\sum_{i\geq 0} -\sum_{i\geq \langle \gamma,\gamma^{(i)}\rangle}\right)\Omega_{n-1-2i}(\gamma)~.
\end{equation}

\subsubsection{Examples of the $2d$-$4d$ Refined Wall-Crossing Formula}
\label{2d4dexamp}

Let us illustrate several basic examples of the refined $2d$-$4d$ wall-crossing formula \eqref{refinedwall2d}. We consider the case where the surface defect has two vacua ($N$=2 above).

\paragraph{Basic Move: $S_{12;\gamma}K_{\gamma'}=K_{\gamma'}S_{12;\gamma+\gamma'}S_{12;\gamma}$}\mbox{}\\

Consider two commuting charges $\gamma$ and $\gamma'$ with $\langle \gamma, \gamma'\rangle =0.$  And further suppose that there is a $4d$ degeneracy along $\gamma'$ with general $4d$ factor $K^{4d}_{\gamma'}$.  Then the following sets of $2d$-$4d$ degeneracies are related by wall-crossing
\begin{equation}
\underbrace{\{\mu_{12}(\gamma,n)=1~,~\omega_{2}(\gamma',j)=-1\}}_{\arg(\mathcal{Z}(\gamma))<\arg(\mathcal{Z}(\gamma'))}\leftrightarrow \underbrace{\{\mu_{12}(\gamma,n)=1~,~\omega_{2}(\gamma',j)=-1~,~\mu_{12}(\gamma+\gamma',n+j)=-1\}}_{\arg(\mathcal{Z}(\gamma'))<\arg(\mathcal{Z}(\gamma))}~. \label{SKKSS}
\end{equation}
Indeed, the left-hand side of \eqref{SKKSS} corresponds to a factorization of $\mathcal{S}^{2d-4d}(q)$ given by (recall that $K_{\gamma'} = K^{4d}_{\gamma'} K^{2d}_{\gamma'}$)
\begin{equation}
\left(\begin{array}{cc}1 & (-1)^{n+1}q^{n/2}X_{\gamma}\\0 &1\end{array}\right)\left(\begin{array}{cc}K^{4d}_{\gamma'} & 0\\0 &K^{4d}_{\gamma'}(1+(-1)^{j+1}q^{j/2}X_{\gamma'})\end{array}\right)~,
\end{equation}
while the right-hand side of \eqref{SKKSS} is
\begin{equation}
\left(\begin{array}{cc}K^{4d}_{\gamma'} & 0\\0 &K^{4d}_{\gamma'}(1+(-1)^{j+1}q^{j/2}X_{\gamma'})\end{array}\right)\left(\begin{array}{cc}1 & (-1)^{n+j}q^{(n+j)/2}X_{\gamma+\gamma'}+(-1)^{n+1}q^{n/2}X_{\gamma}\\0 &1\end{array}\right)~.
\end{equation}
One may verify that these two products are equal. Note that in this wall-crossing formula all the torus algebra variables commute.  Therefore, this simple wall-crossing formula may be viewed as $2d$ wall-crossing decorated by flavors.

A similar example where the quantum torus plays an important role is given as follows.  Consider two non-commuting charges $\gamma$ and $\gamma'$ with $\langle \gamma, \gamma'\rangle =\ell.$  And further suppose that there is a single $4d$ hyperpultiplet along $\gamma'$ so $\Omega_{0}(\gamma')=1$ and the $4d$ factor is $E_{q}(X_{\gamma'}).$  Define a set of $2d$-$4d$ soliton degeneracies by
\begin{equation}
\sum_{m\in \mathbb{Z}}\widetilde{\mu}_{12}(\gamma+k\gamma',m)(-1)^{m}q^{m/2}=(-1)^{n}q^{n/2+k(k-\ell)/2}\binom{\ell}{k}_{q}~, \label{degen}
\end{equation}
where in the above $\binom{\ell}{k}_{q}$ is the $q$-binomial coefficient given by
\begin{equation}
\binom{\ell}{k}_{q}=\begin{cases}\frac{\prod_{j=0}^{k-1}(1-q^{\ell-k+j+1})}{\prod_{i=0}^{k-1}(1-q^{i+1})}&k\leq \ell~, \\ 0 &k>\ell~.\end{cases}
\end{equation}
Then the $2d$-$4d$ degeneracies defined below are related by wall-crossing:
\begin{equation}
\underbrace{\{\mu_{12}(\gamma,n)=1\}}_{\arg(\mathcal{Z}(\gamma))<\arg(\mathcal{Z}(\gamma'))}\leftrightarrow \underbrace{\{\widetilde{\mu}_{12}\}}_{\arg(\mathcal{Z}(\gamma'))<\arg(\mathcal{Z}(\gamma))}~, \label{SKKSS2}
\end{equation}
where the right-hand side of the above denotes the collection of degeneracies defined by \eqref{degen}.
This may be demonstrated using a matrix manipulation similar to the above, and noting 
\begin{equation}
X_{\gamma}E_{q}(X_{\gamma'})=E_{q}(q^{\ell}X_{\gamma'})X_{\gamma}=E_{q}(X_{\gamma'})\prod_{j=0}^{\ell-1}(1+q^{j+1/2}X_{\gamma'})X_{\gamma}~.
\end{equation}

\paragraph{Basic Move: $S_{21; \gamma} K_{\gamma+\gamma'} S_{12;\gamma'}  = S_{12;\gamma'} {\widetilde K}_{\gamma+\gamma'} S_{21;\gamma}$}\mbox{}\\

Again we consider two commuting charges $\gamma$ and $\gamma'$ with $\langle \gamma, \gamma'\rangle =0.$  On two sides of the wall, we assume the following BPS spectra,
\begin{align}
&\underbrace{\{
\mu_{12}(\gamma',1)=-1~,~
\mu_{21}(\gamma,0)=-1~,~
\omega_{2}(\gamma+\gamma',1)=-1\}
}_{\arg(\mathcal{Z}(\gamma))<\arg(\mathcal{Z}(\gamma'))}\notag\\
&\hspace{1.8in}\updownarrow \notag\\
&\underbrace{\{
\mu_{12}(\gamma',1)=-1~,~
\mu_{21}(\gamma,0)=-1~,~
\omega_1(\gamma+\gamma', 1)=-1
\}}_{\arg(\mathcal{Z}(\gamma'))<\arg(\mathcal{Z}(\gamma))}~, \label{SKS}
\end{align}
while the other degeneracies $\mu_{ij}$ and $\omega_i$ are vanishing. 
Note that in contrast to the previous example, here the soliton degeneracies $\mu_{ij}$'s are identical are both sides of the wall, while the 2$d$ BPS particle degeneracy $\omega_i$ jumps. 

 In this case the relevant wall-crossing factors become
 \begin{align}
\begin{split}
&S_{21;\gamma} = \left(\begin{array}{cc}1 & 0 \\ X_\gamma & 1\end{array}\right)\,,~~~~~~~~~~~~~~~~~~~~~~~~~~~~~~
S_{12;\gamma'} = \left(\begin{array}{cc}1 & -q^{1\over2}X_{\gamma'}  \\ 0& 1\end{array}\right)\,,\\
&K_{\gamma'+\gamma}  = 
 \left(\begin{array}{cc} 
 E_q(X_{\gamma'+\gamma}) &0  \\ 0&E_q(qX_{\gamma'+\gamma})\end{array}\right)\,,~~~~~~
 \widetilde K_{\gamma'+\gamma}  = 
 \left(\begin{array}{cc}   E_q(qX_{\gamma'+\gamma}) &0  \\ 0&   E_q(X_{\gamma'+\gamma})\end{array}\right)
\end{split}
\end{align}
On the left-hand side of the basic move we have
\begin{align}
S_{21;\gamma}K_{\gamma'+\gamma} S_{12; \gamma'}
= \left(\begin{array}{cc}   E_q(X_{\gamma'+\gamma}) &0  \\ X_\gamma E_q(X_{\gamma'+\gamma}) &   E_q(qX_{\gamma'+\gamma})\end{array}\right)
 \left(\begin{array}{cc}1 & -q^{1\over2}X_{\gamma'}  \\ 0& 1\end{array}\right)\,.
\end{align}
On the right-hand side  we have
 \begin{align}
S_{12;\gamma'}{\widetilde K}_{\gamma'+\gamma} S_{21; \gamma}
= \left(\begin{array}{cc}   E_q(qX_{\gamma'+\gamma}) &-q^{1\over2} X_{\gamma' }E_q(X_{\gamma'+\gamma})  \\0 &   E_q(X_{\gamma'+\gamma})\end{array}\right)\,
 \left(\begin{array}{cc}1 &0 \\ X_\gamma& 1\end{array}\right)
\end{align}
One may easily verify the wall-crossing formula $S_{21; \gamma} K_{\gamma+\gamma'} S_{12;\gamma'}  = S_{12;\gamma'} {\widetilde K}_{\gamma+\gamma'} S_{21;\gamma}$ is satisfied.

\subsection{$2d$-$4d$ Indices From $2d$-$4d$ BPS States}
\label{sec:2d4dBPS}

In this section we extend the conjectures of  section \ref{sec:CS} to infrared prescriptions for Schur indices of $4d$ $\mathcal{N}=2$ theories in the presence of surface defects, by fusing it with the Cecotti-Vafa formalism of section \ref{sec:CV}.

Since the $2d$-$4d$ wall-crossing invariant is independent of the chamber, it is natural to extract from it  moduli-independent functions such as surface defect Schur indices.  More precisely, we conjecture the following infrared formula:
\begin{equation}\label{IRsurfaceform}
\mathcal{I}_{\mathbb{S}}(q)=(q)_{\infty}^{2r}~\mathrm{Tr}\left[\mathcal{S}^{2d-4d}_{\vartheta, \vartheta+\pi}(q)\mathcal{S}^{2d-4d}_{\vartheta+\pi, \vartheta+2\pi}(q)\right]~.
\end{equation}
Where on the right-hand side $\mathcal{S}_{\vartheta, \vartheta+\pi}(q)$ now denotes the $2d$-$4d$ wall-crossing operator defined in \eqref{2d4dS}-\eqref{refinedwall2d} from the $2d$-$4d$ degenaricies, and the trace operation is the ordinary trace on the $N\times N$ matrix (arising from the defect vacua) as well as the trace operation on the quantum torus algebra defined in \eqref{tracedef}.  

By construction, our IR formula for the defect Schur index is wall-crossing invariant, as it should be.  Furthermore, it reduces to the Cecotti-Vafa formula \eqref{cvindex} for the specialized elliptic genus in the special case where the $4d$ theory is empty, and reduces to the conjecture of \cite{Cordova:2015nma} for the Schur index stated in \eqref{irshur} in the special case where the $2d$ theory is empty.  In section \ref{sec:simex} we will provide strong evidence for our IR formula by matching with the localization result for $\frak{su}(2)$ SYM coupled to the $\mathbb{CP}^{1}$ sigma model defect.  Further applications and checks of our IR formula will be presented in \cite{CGS}.

The physical interpretation for our prescription should be a hybrid of the interpretation for the $4d$ theory and the interpretations of 
the $2d$ formula given in \cite{Gaiotto:2015zna}. In terms of a sphere partition function with a surface defect insertion along the great circle, 
we are just counting configurations of BPS particles and solitons which may contribute to the index, each located 
along the great circle according to the phase of its central charge. 

In terms of an operator counting problem, local operators in the $2d$ story are essentially mapped in the IR to some sort of normal-ordered sequence of soliton-creating and particle-creating operators.  
We are adding to the mix operators creating $4d$ particles. 

In both cases, the matrix structure of the formula keeps track of how solitons or soliton-creating operators interpolate between vacua of the $2d$ defect, 
and the quantum torus structure keeps track of the spin carried by the electromagnetic fields sourced by a sequence of dyonic particles
or operators.

\section{A Relation Between Lines  and Surfaces}
\label{sec:linesurrel}

The $2d$-$4d$ wall-crossing formalism described in the section \ref{sec:2d4dWCF} may be extended to include line defects which lie on the defect.  These may either act as left or right boundary conditions for the surface defect, or yield interfaces between distinct surface defects.  We can use boundary conditions for surface defects and their associated framed $2d$-$4d$ degeneracies to provide an interesting relationship between line defect indices and surface defect indices.

As in the purely $2d$ context of section \ref{pure2dframed}, we consider left and right boundary conditions $B_{\alpha}$ and $B^{\alpha}.$  As always such boundary conditions are characterized by an angle $\vartheta$ specifying the supersymmetry algebra that they preserve.  In the infrared on the Coulomb branch these boundary conditions support framed BPS states which may now carry $4d$ gauge charges $\gamma$, and we promote the indices accordingly
\begin{equation}
\chi^{\alpha, i}(X_{\gamma},q)= \sum_{\gamma\in \Gamma}\mathrm{Tr}_{i,B^{\alpha}(\vartheta),\gamma}\left((-1)^{F_{2d}+2C}q^{C}\right)X_{\gamma}~, \hspace{.15in}\chi_{\alpha, i}(X_{\gamma},q)= \sum_{\gamma\in \Gamma}\mathrm{Tr}_{B^{\alpha},(\vartheta),i,\gamma}\left((-1)^{F_{2d}+2C}q^{C}\right)X_{\gamma}~.
\end{equation}
We assemble these degeneracies into $N$-component row vectors $F(B_{\alpha},\vartheta,X_{\gamma},q)$ and $N$-component column vectors $F(B^{\alpha},\vartheta,X_{\gamma},q)$.  

We have several structures analogous  to the pure 2$d$ case, with a twist. For example, if we bring together a left and a right boundary condition for a surface defect, the result is a line defect in the 4$d$ theory, whose framed BPS degeneracies are 
the inner product 
\begin{equation}
\left(B_{\alpha}, B^{\beta}\right) = F(B_{\alpha},\vartheta,X_{\gamma},q)F(B^{\beta},\vartheta,X_{\gamma},q)\,.
\end{equation}

As in the purely two-dimensional context, these framed $2d$-$4d$ degeneracies may jump as the supersymmetry angle $\vartheta$ is changed.  This is governed a formula which intertwines the $2d$ framed wall-crossing formula \eqref{2dframedwc} with the $4d$ framed wall-crossing formula \eqref{4dframedwc}.  This takes the following form:
\begin{equation}
\mathcal{S}^{2d-4d}_{\vartheta', \vartheta}F(B^{\alpha},\vartheta, X_{\gamma},q)=F(B^{\alpha},\vartheta', X_{\gamma},q)\mathcal{S}^{4d}_{\vartheta', \vartheta}~, \hspace{.25in}F(B_{\alpha},\vartheta, X_{\gamma},q)\mathcal{S}^{2d-4d}_{\vartheta, \vartheta'}=\mathcal{S}^{4d}_{\vartheta, \vartheta'}F(B_{\alpha},\vartheta', X_{\gamma},q)~.\label{2d4dframedwc}
\end{equation}

As in $2d$, we can write a generalization of our index formula which describes local operators at the tip of a 
wedge formed by a surface defect bounded by two boundary conditions: 
\begin{equation}\label{IRsurfacewedge}
\mathcal{I}_{B_\alpha(\vartheta), B^\beta(\vartheta')}(q)=(q)_{\infty}^{2r}\mathrm{Tr}\left[ F(B_{\alpha},\vartheta,X_{\gamma},q) \mathcal{S}^{2d-4d}_{\vartheta, \vartheta'}(q)F(B^{\beta},\vartheta,X_{\gamma},q)\mathcal{S}^{4d}_{\vartheta', \vartheta+2\pi}(q)\right]~,
\end{equation}
where the trace should be defined with care by splitting the factors into two blocks associated to $[\vartheta,\vartheta + \pi]$ and 
$[\vartheta + \pi, \vartheta + 2 \pi]$, as for bulk line defects. 

We again have a monodromy operator $\mathcal{R}_{2\pi}$ which results from parallel transporting the supersymmetry angle from $\vartheta$ to $\vartheta+2\pi$. Given a sufficiently complete basis of boundary conditions, we can repeat calculations analogous to 
the 2$d$ case: introduce an identity interface in the bare 2$d$-4$d$ index, decompose its framed BPS degeneracies 
as a bi-linear of $F(B_{\alpha},\vartheta,X_{\gamma},q)$ and $F(B^{\beta},\vartheta,X_{\gamma},q)$,
and transport $B_{\alpha}$ around a $2 \pi$ angle to ``unwrap'' the surface defect. 

The result of the manipulation is a 4$d$ Schur index with the insertion of a Laurent polynomial $\mathcal{ O}^{2d}(X,q)$ in the $X_\gamma$ 
built from $\left(B_{\alpha}, B^{\beta}\right)$ and $\left(\mathcal{R}_{2\pi} \circ B_{\alpha}, B^{\beta}\right)$.
It is clear that the final expression $\mathcal{ O}^{2d}(X,q)$ does not to depend on the basis of line defects employed 
in the calculation. Indeed, the framed BPS degeneracies of the defects themselves drop out of the calculation and 
$\mathcal{ O}^{2d}(X,q)$ can be computed by the following simple algorithm:
\begin{itemize}
\item Move the quantum dilogarithms in $\mathcal{S}^{2d-4d}_{\vartheta, \vartheta+\pi}(q)$ all the way to the right, 
leaving an overall $N \times N$ matrix $S_+(X,q)$ on the left: 
\begin{equation}
\mathcal{S}^{2d-4d}_{\vartheta, \vartheta+\pi}(q) = S_+(X,q) \mathcal{S}^{4d}_{\vartheta, \vartheta+\pi}(q)\,.
\end{equation}
\item Move the quantum dilogarithms in $\mathcal{S}^{2d-4d}_{\vartheta+\pi, \vartheta+2\pi}(q)$ all the way to the left, 
leaving an overall $N \times N$ matrix $S_-(X,q)$ on the right: 
\begin{equation}
\mathcal{S}^{2d-4d}_{\vartheta+\pi, \vartheta+2\pi}(q) = \mathcal{S}^{4d}_{\vartheta+\pi, \vartheta+2\pi}(q) S_-(X,q) \,.
\end{equation}
\item Take the $N$-by-$N$ matrix trace  (but not the trace in the quantum torus algebra)
\begin{equation}
\mathcal{ O}^{2d}(X,q) = \mathrm{Tr}_{N \times N}S_-(X,q)  S_+(X,q)\,.
\end{equation}
\end{itemize} 
The result $\mathcal{ O}^{2d}(X,q)$ is the framed BPS degeneracy of whatever 4$d$ line defect arises from unwrapping the 
surface defect. 

We can expand the insertion in a basis of line defects $L_{j}$, and their associated framed BPS generating functions
\begin{equation}
\mathcal{O}^{2d}=\sum_{j}c_{j}(q)F(L_{j},\vartheta,X_{\gamma},q)~.
\end{equation}
The coefficients $c_{j}(q)$ appearing above are functions only of $q$, and do not depend on the quantum torus variables.  
We therefore derive the following non-trivial relationship between surface defect indices and line defect indices
\begin{equation}
\mathcal{I}_{\mathbb{S}}(q) =\sum_{j}c_{j}(q)\mathcal{I}_{L_{j}}(q)~. \label{linesurfrel}
\end{equation}
As we will see in section \ref{sec:simex}, this equation is practically calculable in examples.

We can describe the physical picture behind the relationship \eqref{linesurfrel} in the language of a sphere partition function as well. 
Consider the defect indices as partition functions on $S^{3}\times S^{1}$.  Initially we have a surface defect wrapping an equatorial circle times $S^{1}$.  The insertion of a resolution of the identity corresponds to cutting the surface defect open using boundary conditions $(B_{\alpha},\vartheta),$  $(B^{\alpha},\vartheta).$  

An important aspect of this picture is that while in flat space the angle $\vartheta$ specifies the supersymmetry algebra preserved by the boundary condition, on the sphere it has a more geometric meaning: it is simply the angular position along the equator.  In particular, this means that the process of parallel transporting $B_{\alpha_{i}}$ from $\vartheta$ to $\vartheta+2\pi$ is literally unwrapping the surface defect along the equator.  At the end of the process the surface defect is gone, but what remains is a sum of lines.  This is illustrated in Figure \ref{fig:cut}.

Finally, let us comment on some implications of this result that are particular to the conformally invariant case.  As mentioned in section \ref{sec:schur}, for a conformal field theory the Schur index $\mathcal{I}(q)$ is known abstractly to be the vacuum character of an associated $2d$ chiral algebra \cite{Beem:2013sza}.  Moreover, for a conformally invariant surface defect, the index $\mathcal{I}_{\mathbb{S}}(q)$ is a (not necessarily vacuum) character of the same chiral algebra \cite{CGS,BPR}.  Therefore we deduce from \eqref{linesurfrel} that the sum of line defect indices that results from unwrapping such a surface is also a character.  Frequently, it is possible to invert such relationships and thus conclude that the individual line defect indices are themselves linear combinations of characters of chiral algebras.  In particular this explains the surprising observations of \cite{Cordova:2016uwk} regarding such line defect indices.   We will demonstrate this in the case of Argyres-Douglas theory in \cite{CGS}.

\section{$\frak{su}(2)$ SYM Coupled to the $\mathbb{CP}^{1}$ Sigma Model }
\label{sec:simex}

In this section we discuss the application of the infrared formula \eqref{IRsurfaceform} in the example of the $\mathbb{CP}^{1}$ sigma model coupled to $\frak{su}(2)$ SYM.  In the context of class $S$ constructions, this is the canonical surface defect for this $4d$ theory. The Schur index was computed in \eqref{cp1surface} using supersymmetric localization of the Lagrangian description of this defect.  The resulting index is repeated here for convenience
\begin{equation}
\mathcal{I}_{\mathbb{S}}(q)=1+2\sum_{n=1}^{\infty}q^{n^{2}}=\theta_{3}(2\tau)~. \label{su2examprepeat}
\end{equation}

We now aim to reproduce this result using the $2d$-$4d$ BPS spectrum and our conjectured infrared formula \eqref{IRsurfaceform}.  We work in the strongly-coupled chamber of the bulk $4d$ system where there are only two $4d$ BPS particles, the monopole and the dyon.  We denote the electromagnetic charge vectors for these states as $\gamma$ and $\gamma'$ with Dirac pairing $\langle \gamma,\gamma'\rangle=2$.  

The presence of the surface defect further divides the strongly-coupled chamber into several subchambers with different $2d$-$4d$ BPS spectra.  We carry out our calculation in the $L$ chamber in \cite{Gaiotto:2011tf} where there are two 4$d$ particles,  two 2$d$ particles,   and two $2d$ solitons interpolating between the two vacua. The two 2$d$ particles  carry 4$d$ charge $\gamma$ and $\gamma'$, respectively, and they both have unit $C$-charge and live in the first vacuum with degeneracy $\omega_1( \gamma , 1) = -1$ and $\omega_1(\gamma',1)=-1$.  The 4$d$ charges of the two 2$d$ solitons are 0 and $\gamma$, respectively, with degeneracies $\mu_{12}(0,0)=1$ and $\mu_{21}(\gamma, 0)=1$.  In addition to the above 2$d$-4$d$ BPS states, we of course also have their antiparticles.

 The $2d$-$4d$ BPS states in this chamber in increasing phase order are,
\begin{align}
\gamma_{12}+\gamma \,, \gamma\,, \gamma_{21}\,,\gamma' \,,
\end{align}
where $\gamma$ denotes collectively the 4$d$ particle and the 2$d$ particle carrying charge $\gamma$, and similarly for $\gamma'$.  The corresponding wall-crossing factors are
\begin{equation}
S_{12;\gamma } =  \begin{pmatrix}1 & -q^{1\over2} X_\gamma \cr 0 & 1\end{pmatrix} \qquad \qquad S_{21;0} =  \begin{pmatrix}1 & 0 \cr 1 & 1\end{pmatrix}
\end{equation}
and 
\begin{equation}
K_\gamma=  \begin{pmatrix}E_q(qX_\gamma) & 0 \cr 0 & E_q(X_\gamma)\end{pmatrix} \qquad \qquad K_{\gamma'}=  \begin{pmatrix}E_q( qX_{\gamma'}) & 0 \cr 0 & E_q(X_{\gamma'})\end{pmatrix} \,.
\end{equation}
Here $K_\gamma$ is the product of the $K$-factors from the 2$d$ and the 4$d$ particles, $K_\gamma =  K^{2d}_\gamma  K^{4d}_\gamma$.   Similar expressions hold for the wall-crossing factors of the antiparticles with central charge phases between $\vartheta+\pi$ to $\vartheta+2\pi$.  Taking the product of the wall-crossing factors for the particles in this chamber, we obtain the following explicit formula for the spectrum generator:
\begin{align}
\mathcal{S}^{2d-4d}_{\vartheta, \vartheta+\pi}=
&S_{12;\gamma} K_\gamma S_{21;0}  K_{\gamma'}
=\left(\begin{array}{cc}
E_q(X_\gamma) E_q(qX_{\gamma'})
 & -q^{1\over2} X_{\gamma} E_q(X_\gamma) E_q(X_{\gamma'}) \\
 E_q(X_\gamma) E_q(qX_{\gamma'}) &
  E_q(X_\gamma) E_q(X_{\gamma'})\end{array}\right)\,,
\end{align}
with a similar expression for $\mathcal{S}^{2d-4d}_{\vartheta+\pi, \vartheta+2\pi}.$  

Using our IR formula \eqref{IRsurfaceform} for the surface defect index, we have
\begin{align}
\mathcal{I}_\mathbb{S}(q)=(q)_\infty^2 \text{Tr}\left[ 
S_{12; \gamma} K_\gamma S_{21;0 } K_{\gamma'}
S_{21; - \gamma} K_{-\gamma} S_{12;0} K_{-\gamma'}
\right]\,.
\end{align} 
Taking the matrix trace first, the surface defect index becomes:
\begin{align}\label{pure}
&(q)_\infty^2 \text{Tr}\left[
E_q(X_{\gamma})E_q(qX_{\gamma'})E_q(q^{-1}X_{-\gamma})E_q(q^{-1}X_{-\gamma'})\right.\notag\\
&\left.
-(q^{1\over2} X_\gamma) E_q(X_{\gamma})E_q(X_{\gamma'})
(q^{-\frac12} X_{-\gamma})
E_q(q^{-{1}}X_{-\gamma})E_q(q^{-{1}}X_{-\gamma'})\right.\notag\\
&\left.
-  E_q(X_{\gamma})E_q(qX_{\gamma'})
E_q(q^{-1}X_{-\gamma})E_q(X_{-\gamma'})
+E_q(X_{\gamma})E_q(X_{\gamma'})E_q(q^{-{1}}X_{-\gamma})E_q(X_{-\gamma'})\right]\,.
\end{align}
To proceed, we replace each factor $q^{1\over2}X_\gamma$ that appears outside a dilogarithm by $-1+(1+q^{1\over2}X_\gamma)$, and then use $(1+q^{1\over2}X_\gamma)E_q(X_\gamma)=E_q(qX_\gamma)$ to get eliminate all such factors. We also perform a similar manipulation for $q^{-\frac12}X_{-\gamma}$.  After this replacement, we can rewrite \eqref{pure} as
\begin{align}\label{SU(2)IR}
& \mathcal{I}_\mathbb{S}(q)=(q)_\infty^2 \sum_{\ell_1,\ell_2=0}^\infty {q^{\ell_1+\ell_2+2\ell_1\ell_2} \over (q)_{\ell_1}^2 (q)_{\ell_2}^2} \left(
2q^{-\ell_2} - q^{\ell_1-\ell_2} -q^{-\ell_1-\ell_2} +2q^{-\ell_1} -q^{-\ell_1+\ell_2}
\right)\notag\\
&=1+2q+2q^4+2q^9+2q^{16}+2q^{25}+\cdots\,. 
\end{align}
Note that the final answer above is an insertion $\left(
2q^{-\ell_2} - q^{\ell_1-\ell_2} -q^{-\ell_1-\ell_2} +2q^{-\ell_1} -q^{-\ell_1+\ell_2}
\right)$ into the double-sum formula for the Schur index $\mathcal{I}(q)$ of the theory without the defect, which is \cite{Cordova:2015nma}:
\begin{align}
\mathcal{I}(q) =(q)_\infty^2 \sum_{\ell_1,\ell_2=0}^\infty {q^{\ell_1+\ell_2+2\ell_1\ell_2} \over (q)_{\ell_1}^2 (q)_{\ell_2}^2} = 1+q^2+q^6+q^{12}+q^{20}+q^{30}+q^{42}+\cdots
={q^{-\frac18}\over 2}\theta_2(2\tau)\,.
\end{align}
We have checked the above IR formula \eqref{SU(2)IR} for the canonical surface defect index agrees with the localization answer \eqref{su2examprepeat} to $\mathcal{O}(q^{125})$.

\subsection{Resolving the Surface into Lines}

In the example of $\frak{su}(2)$ SYM coupled to $\mathbb{CP}^1$ sigma model,  both the 4$d$ bulk theory and the $2d$ theory living on the surface defect have Lagrangian descriptions, and the $2d$-$4d$ index can be computed using localization as demonstrated in section \ref{sec:2d4dindices}.  There the final expression \eqref{su2localization} (see also \eqref{CP1special}) takes a very suggestive form: 
\begin{align}
\mathcal{I}_\mathbb{S}(q)  = \mathcal{I}(q)  -  \mathcal{I}_{L_\mathbf{3}}(q)\,,
\end{align}
where $\mathcal{I}(q)$ and $\mathcal{I}_{L_\mathbf{3}}(q)$  are respectively, the index without defects and the index with a triplet half Wilson line defect.  The latter has been computed both using the localization method and from framed BPS states in \cite{Cordova:2016uwk}.  The above relation indicates that, for the purpose of certain supersymmetric computations, the canonical surface defect $\mathbb{S}$ can be decomposed into a trivial an a triplet Wilson line defect $L_\mathbf{3}$,  as explained in section \ref{sec:linesurrel}. In this section we will explicitly demonstrate this phenomenon in the example of $\frak{su}(2)$ SYM using our infrared formula.

We will follow the general algorithm in section \ref{sec:linesurrel} to rearrange the spectrum generator into a form where all the $S$-factors are multiplied together.  This can be done by repeatedly applying the first basic wall-crossing formula $S_{12;\gamma} K_{\tilde \gamma}  =  K_{\tilde\gamma} S_{12;\gamma+\tilde \gamma} S_{12;\gamma}$ in section \ref{2d4dexamp}.  The trace of the quantum spectrum generator can be rewritten as
\begin{align}
&\text{Tr} \left[ \mathcal{S}_{\vartheta ,\vartheta +\pi } \mathcal{S}_{\vartheta+\pi ,\vartheta+2\pi}\right]
=\text{Tr}\left[
S_{12;\gamma} K_\gamma S_{21;0} K_{\gamma'}\,
S_{21;-\gamma} K_{-\gamma} S_{12;0} K_{-\gamma'}
\right]\notag\\
&=\text{Tr}\left[
K_{-\gamma'}  S_{12;\gamma} K_\gamma K_{\gamma'} S_{21;\gamma'} S_{21;0}
S_{21;-\gamma} K_{-\gamma}  S_{12;0}
\right]\notag\\
&=\text{Tr}\left[
S_{12;\gamma} S_{12;\gamma-\gamma'} K_{-\gamma'}  K_\gamma K_{\gamma'} S_{21;\gamma'} S_{21;0}
S_{21;-\gamma} K_{-\gamma}  S_{12;0}
\right]\notag\\
&= \text{Tr} \left[
S_{12;\gamma} K_{-\gamma} S_{12;\gamma-\gamma'} K_{-\gamma'}  K_\gamma K_{\gamma'} S_{21;\gamma'} S_{21;0}
S_{21;-\gamma}
\right]\notag\\
&=\text{Tr}  \left[ 
\Sigma^{2d}\, K_{-\gamma} K_{-\gamma'}K_\gamma K_{\gamma'}
\right]\,,
\end{align}
where $\Sigma^{2d}$ is a two-by-two matrix defined as
\begin{align}
\Sigma^{2d}&= S_{21;\gamma'} S_{21;0}
S_{21;-\gamma}
S_{12;\gamma} S_{12;\gamma-\gamma'} S_{12;-\gamma'}\notag\\
&= \left(\begin{array}{cc}1 & 0 \\1+q^{1\over2} X_{\gamma'} +q^{-\frac12} X_{-\gamma} & 1\end{array}\right)
\left(\begin{array}{cc}1 & -X_{\gamma-\gamma'}-q^{1\over2} X_{\gamma} -q^{-\frac12} X_{-\gamma'}  \\0 & 1\end{array}\right)\notag\\
&= \left(\begin{array}{cc}1 &  -X_{\gamma-\gamma'}-q^{1\over2} X_{\gamma} -q^{-\frac12} X_{-\gamma'} 
 \\1+q^{1\over2} X_{\gamma'} +q^{-\frac12} X_{-\gamma} & -F(L_\mathbf{3})\end{array}\right)\,.
\end{align}
In the last line $F(L_\mathbf{3})$ is the generating function for a triplet half Wilson line defect \cite{Cordova:2013bza},
\begin{align}
F(L_\mathbf{3})  =(X_{\gamma+\gamma'} +1 + X_{-\gamma-\gamma'})+
(q^{\frac12} +q^{-\frac12})(X_\gamma +X_{-\gamma'})+X_{\gamma-\gamma'}\,.
\end{align}


Now our infrared formula for the surface defect index can be written as that for the $4d$ Schur index with the insertion of $\mathcal{O}^{2d}   = \text{Tr}_{2\times 2} \Sigma^{2d}  =  1- F(L_{\bf3})$,
\begin{align}
\mathcal{I}_\mathbb{S} &=
  (q)_\infty^2\text{Tr}\left[ \, \left( 1- F(L_\mathbf{3}) \right) \,E_q(X_{-\gamma})E_q(X_{-\gamma'}) E_q(X_\gamma) E_q(X_{\gamma'})\right]\notag\\
&= \mathcal{I}(q) - \mathcal{I}_{L_\mathbf{3}}(q)\,,
\end{align}
where in the last line we have used the infrared formulas for the Schur index in the presence of line defects \cite{Cordova:2016uwk} discussed in section \ref{sec:linedef}.

To sum up, in this section we showed that our infrared formula for the surface defect index \eqref{IRsurfaceform} not only reproduces the correct answer from localization techniques, but also  confirms the physical picture that a surface defect can be decomposed into multiple lines as anticipated in section \ref{sec:linesurrel}.

 \section*{Acknowledgements} 
We thank Tomoyuki Arakawa, Chris Beem, Thomas Creutzig, Leonardo Rastelli, and Andy Neitzke for interesting discussions.  We are grateful to the workshop ``Exact Operator Algebras in Superconformal Field Theories" supported by the Simons Foundation for providing a stimulating research environment.  
The work of CC is supported by a Martin and Helen Chooljian membership at the Institute for Advanced Study and DOE grant DE-SC0009988.  The research of DG was supported by the Perimeter Institute for Theoretical Physics.  Research at Perimeter Institute is supported by the Government of Canada through Industry Canada and by the Province of Ontario through the Ministry of Economic Development \& Innovation.  SHS is supported by the National Science Foundation grant PHY-1606531.

\appendix

\section{Supercharges Preserved by Surface Defects}
\label{chargedetails}

In this paper we  consider both conformal and asymptotically free 4$d$ $\mathcal{N}=2$ theories and their surface defects.  Here we identify the symmetry algebra preserved by a surface defect in a conformal theory, while the non-conformal case can be trivially carried over by restricting to the supersymmetric subalgebra. We also show that in the conformal case, the  supercharges used to define the chiral algebra are preserved by the surface defect.

 Let us set up our convention on the 4$d$ $\mathcal{N}=2$ superconformal algebra $\frak{su}(2,2|2)$. Its bosonic subgroup is $\frak{so}(2,4)\times \frak{su}(2)_R\times \frak{u}(1)_r$, where $\frak{so}(2,4)$ is the four-dimensional bosonic conformal algebra while $\frak{su}(2)_R\times \frak{u}(1)_r$ is the $R$-symmetry.  We will use $A,B =1,2$ to denote the doublet index of  $\frak{su}(2)_R$, and $\alpha,\beta= +,-$, $\dot\alpha,\dot\beta= \dot +,\dot -$   to denote the doublet indices of $\frak{su}(2)_{1}\times \frak{su}(2)_2 =\frak{so}(4)_{\text{rotation}}$.  All the doublet indices will be raised and lowered by the antisymmetric symbol $\epsilon^{AB}$ with $\epsilon^{12}=\epsilon_{21}=+1$.
 The nontrivial anticommutators among the sixteen fermionic generators $\{Q^A_{~\alpha},~ \tilde Q_{A\dot \alpha},~S_A^{~\alpha}, ~\tilde S^{A\dot \alpha}\}$ are
\begin{align}
\begin{split}
&\{Q^A_{~\alpha}  ,  \tilde Q_{B\dot \beta} \} =2 \delta^A_B \sigma^\mu_{\alpha \dot \beta} P_\mu 
= \delta^A_B P_{\alpha \dot \beta}
\,,\\
&\{\tilde S^{A\dot \alpha} , S_B^{~\beta} \} =2 \delta^A_B \bar\sigma^{\mu\dot\alpha\beta} K_\mu  
=  \delta^A_B  K^{\dot \alpha \beta}
\,,\\
&\{Q^A_{~\alpha} , S_B^{~\beta} \} = {1\over 2} \delta^A_B \delta^\beta_\alpha \Delta +\delta^A_B M_\alpha^{~\beta} - \delta_\alpha^\beta R^A_{~B}\, ,\\
&\{\tilde S^{A\dot \alpha} ,\tilde Q_{B\dot \beta} \} ={1\over 2}\delta^A_B \delta^{\dot \alpha}_{\dot \beta} \Delta 
+\delta^A_B M^{\dot \alpha}_{~\dot \beta} +\delta^{\dot \alpha}_{\dot\beta} R^A_{~B}\,.
\end{split}
\end{align}
Here $\Delta$ is the dilation generator and $M_\alpha^{~\beta}, M^{\dot\alpha}_{~\dot \beta}$ are the $\frak{so}(4)_{\text{rotation}}$ rotation generators satisfying $M_\alpha^{~\alpha}=M^{\dot \alpha}_{~\dot\alpha}=0$.  
 $R^A_{~B}$ contains the $\frak{su}(2)_R$ generators $R^\pm, R$ and the $\frak{u}(1)_r$ generator $r$,
\begin{align}
R^1_{~1} = {1\over 2}r -R \, ,~~~R^1_{~2}  = R^-\, ,~~~R^2_{~1}  = R^+\, ,~~~ R^2_{~2} = {1\over 2}r +R\,,
\end{align}
where $[R,R^\pm ] = \pm R^\pm$ and $[R^+,R^-] = 2R$.\footnote{In this convention a lower $\frak{su}(2)_R$ doublet index $A=1$ and $A=2$ have $+1/2$ and $-1/2$ eigenvalue under $R$, respectively.}

A surface defect $\mathbb{S}$  preserves an  $\frak{su}(1,1|1)\times \frak{su}(1,1|1)\times \frak{u}(1)_C$ subalgebra of $\frak{su}(2,2|2)$. Note that $\frak{su}(1,1|1)\times \frak{su}(1,1|1)$ is the global part of the 2$d$ (2,2) NS-NS superconformal algebra and $\frak{u}(1)_C$ is the commutant of embedding.  The  $\frak{su}(1,1|1)\times \frak{su}(1,1|1)$ consists of the generators $L_{0,\pm1}$ and $\bar L_{0,\pm 1}$ of global bosonic conformal algebra $\frak{sl}(2,\mathbb{R})\times \frak{sl}(2,\mathbb{R})$, the generators $J_0,\bar J_0$ of the $\frak{u}(1)_L\times \frak{u}(1)_R$ $R$-symmetry, and  four supercharges $G^\pm_{- {1\over 2}},\bar G^\pm _{-{\frac12}}$ as well as four superconformal fermionic generators $G^\pm_{+{1\over 2}},\bar G^\pm _{+{\frac12}}$.\footnote{In this Appendix we use different notations for the supercharges in the 2$d$ (2,2) supersymmetry algebra compared to \eqref{qcharges}. The two set of notations are related by $G^+_{-\frac12}  =Q_+$, $G^-_{-\frac12}=\bar Q_+$, $\bar G^+_{-\frac12} = Q_-$, $\bar G^-_{-\frac12}  = \bar Q_-$.  } 
 The nonzero (anti)commutators  are
\begin{align}
&[L_0, G^\pm_r] = -r G^\pm_r\,,~~~~~~~[\bar L_0,\bar G^\pm_r] = -r \bar G^\pm_r\,,\notag\\
&[J_0,G^\pm_r] =\pm G^\pm_r\,,~~~~~~~~~[\bar J_0, \bar G^\pm_r]=\pm \bar G^\pm_r\,,\notag\\
&\{ G^+_r ,G^-_s\}=  L_{r+s}  +{ r-s \over 2}J_{r+s} \,,~~~~\{ \bar G^+_r ,\bar G^-_s\}= \bar L_{r+s}  +{r-s\over2} \bar J_{r+s}\,,~~~~~r,s=\pm\frac12\,.
\end{align}

Let us pick a convention for the embedding of $\frak{su}(1,1|1)\times \frak{su}(1,1|1)\times \frak{u}(1)_C$ into $\frak{su}(2,2|2)$.  We will orient the surface defect to be along the 12-plane.  We will choose 
\begin{align}
M_{\perp}\equiv M_+^{~+} +M^{\dot +}_{~\dot +}
\end{align}
 to be the rotation on the 34-plane and
\begin{align}
M_{||}\equiv M_+^{~+} -M^{\dot +}_{~\dot +}
\end{align}
to be the rotation on the 12-plane where the surface defect $\mathbb{S}$ lies on.  
The supercharges of $\frak{su}(1,1|1)\times \frak{su}(1,1|1)$ are identified as
\begin{align}
G^+_{-\frac12} = Q^2_{~+}\,,~~~~~G^-_{-\frac12} = \tilde Q_{2\dot -}\,,~~~~~
\bar G^+_{-\frac12}  =Q^1_{~-}\,,~~~~~\bar G^-_{-\frac12} = \tilde Q_{1\dot +}\,,
\end{align}
and similarly for their superconformal counterparts,
\begin{align}
G^+_{+\frac12} =  \tilde S^{2\dot -}\,,~~~~~G^-_{+\frac12} =S_2^{~+}\,,~~~~~
\bar G^+_{+\frac12}  =\tilde S^{1\dot +}\,,~~~~~\bar G^-_{+\frac12} =S_1^{~-} \,.
\end{align}
From the anticommutators between the above supercharges, we  identify 
\begin{align}
\begin{split}\label{embed1}
&L_0=\frac12 (\Delta+M_{||}) \,,~~~\,~~~~~~~~\bar L_0=\frac12 (\Delta-M_{||})\,,\\
&J_0=  2R-M_{\perp} + r\,,~~~~~~~~~~\bar J_0 = - 2R + M_{\perp}+r\,.
\end{split}
\end{align}
Finally, the commutant $\frak{u}(1)_C$ is generated by 
\begin{align}\label{embed2}
C=R-M_{\perp}\,.
\end{align}

Incidentally, the four supercharges $\mathtt{Q}_i$ and $\mathtt{Q}_i^\dagger$ $(i=1,2)$ that are used to construct the chiral algebra in \cite{Beem:2013sza} are preserved 
 by the $\frak{su}(1,1|1)\times \frak{su}(1,1|1)\times \frak{u}(1)_C$ subalgebra of the surface defect. More explicitly, 
\begin{align}
\begin{split}\label{ChiralAlgebraQ}
&\mathtt{Q}_1 \equiv Q^1_{~-} + \tilde S^{2\dot -}\,,~~~\mathtt{Q}_2 \equiv  S_1^{~-}- \tilde Q_{2\dot -}\,,\\
&\mathtt{Q}_1^\dagger  \equiv  S_1^{~-}+ \tilde Q_{2\dot -}\,,~~~\mathtt{Q}_2^\dagger \equiv Q^1_{~-} - \tilde S^{2\dot -}\,.
\end{split}
\end{align}
  Note that in this convention the chiral algebra lies on the 34-plane, which is transverse to the surface defect.

The 4$d$ $\mathcal{N}=2$ superconformal index can be defined as 
\begin{align}
\mathcal{I}(q,p,t) = \text{Tr} (-1)^{F_{4d}} q^{-M_{\perp}-r} p^{M_{||}-r} t^{R+r}\,,
\end{align}
where the exponents  of the fugacities $q,p,t$ are the maximal set of quantum numbers that commute with a particular supercharge, which is chosen to be $\tilde Q_{1\dot +}$ here.  We have chosen the 4$d$ fermion number to be $(-1)^{F_{4d}} = e^{2\pi i R}$ \eqref{4dfermion}. The index only receives contribution from operators satisfying $\{\tilde Q_{1\dot+} ,\tilde S^{1\dot+}\} = \Delta +M_{\perp}-M_{||}-2R+r=0$.

The 2$d$ NS-NS elliptic genus, on the other hand, is defined as
\begin{align}
\mathcal{G}(\mathbf{q}, y,e)  = \text{Tr}_{NSNS}\left[\, (-1)^{F_{2d}} \mathbf{q}^{L_0} \mathbf{\bar q}^{\bar L_0 - \bar J_0} y^{J_0} e^{C}\,\right]\,,
\end{align}
where we have introduced a fugacity $e$ for the commutant $\frak{u}(1)_C$, which is a flavor symmetry from the 2$d$ point of view.  $F_{2d}=J_0+\bar J_0$ is the 2$d$ fermion number.  The elliptic genus received contribution from operators that are annihilated by $\bar G_{-\frac12}^-$, which is $\tilde Q_{1\dot +}$ when embedded into $\frak{su}(2,2|2)$.

Using \eqref{embed1} and \eqref{embed2}, we find that the 4$d$ fugacities $q,p,t$ are related to the 2$d$ fugacities  $\mathbf{q},y, e$ as
\begin{align}
\mathbf{q}=p\,,~~~~~y= q^{-1}p^{-\frac12}t\,,~~~~~e=q^2t^{-1}
\end{align}

The Schur limit of the 4$d$ index is $t=q$ and the $p$ dependence will drop out due to enhanced supersymmetry.  This translates into the following limit on the 2$d$ fugacities,
\begin{align}
\mathbf{q}\, y^2=1\,.
\end{align}

\bibliography{2d4ddraft}{}

\providecommand{\href}[2]{#2}\begingroup\raggedright\begin{thebibliography}{10}

\bibitem{Gukov:2014gja}
S.~Gukov, ``{Surface Operators},'' in {\em New Dualities of Supersymmetric
  Gauge Theories}, J.~Teschner, ed., pp.~223--259.
\newblock 2016.
\newblock
\href{http://www.arXiv.org/abs/1412.7127}{{\tt 1412.7127}}.
\newblock

\bibitem{Gaiotto:2013sma}
D.~Gaiotto, S.~Gukov, and N.~Seiberg, ``{Surface Defects and Resolvents},''
  {\em JHEP} {\bf 09} (2013) 070,
\href{http://www.arXiv.org/abs/1307.2578}{{\tt 1307.2578}}.

\bibitem{Gukov:2006jk}
S.~Gukov and E.~Witten, ``{Gauge Theory, Ramification, And The Geometric
  Langlands Program},''
\href{http://www.arXiv.org/abs/hep-th/0612073}{{\tt hep-th/0612073}}.

\bibitem{Gukov:2008sn}
S.~Gukov and E.~Witten, ``{Rigid Surface Operators},'' {\em Adv. Theor. Math.
  Phys.} {\bf 14} (2010), no.~1, 87--178,
\href{http://www.arXiv.org/abs/0804.1561}{{\tt 0804.1561}}.

\bibitem{Gomis:2007fi}
J.~Gomis and S.~Matsuura, ``{Bubbling surface operators and S-duality},'' {\em
  JHEP} {\bf 06} (2007) 025,
\href{http://www.arXiv.org/abs/0704.1657}{{\tt 0704.1657}}.

\bibitem{Hanany:1997vm}
A.~Hanany and K.~Hori, ``{Branes and N=2 theories in two-dimensions},'' {\em
  Nucl. Phys.} {\bf B513} (1998) 119--174,
\href{http://www.arXiv.org/abs/hep-th/9707192}{{\tt hep-th/9707192}}.

\bibitem{Gaiotto:2009fs}
D.~Gaiotto, ``{Surface Operators in N = 2 4d Gauge Theories},'' {\em JHEP} {\bf
  11} (2012) 090,
\href{http://www.arXiv.org/abs/0911.1316}{{\tt 0911.1316}}.

\bibitem{Gaiotto:2011tf}
D.~Gaiotto, G.~W. Moore, and A.~Neitzke, ``{Wall-Crossing in Coupled 2d-4d
  Systems},'' {\em JHEP} {\bf 12} (2012) 082,
\href{http://www.arXiv.org/abs/1103.2598}{{\tt 1103.2598}}.

\bibitem{Nakayama:2011pa}
Y.~Nakayama, ``{4D and 2D superconformal index with surface operator},'' {\em
  JHEP} {\bf 08} (2011) 084,
\href{http://www.arXiv.org/abs/1105.4883}{{\tt 1105.4883}}.

\bibitem{Gadde:2013ftv}
A.~Gadde and S.~Gukov, ``{2d Index and Surface operators},'' {\em JHEP} {\bf
  03} (2014) 080,
\href{http://www.arXiv.org/abs/1305.0266}{{\tt 1305.0266}}.

\bibitem{Gaiotto:2012xa}
D.~Gaiotto, L.~Rastelli, and S.~S. Razamat, ``{Bootstrapping the superconformal
  index with surface defects},'' {\em JHEP} {\bf 01} (2013) 022,
\href{http://www.arXiv.org/abs/1207.3577}{{\tt 1207.3577}}.

\bibitem{Alday:2013kda}
L.~F. Alday, M.~Bullimore, M.~Fluder, and L.~Hollands, ``{Surface defects, the
  superconformal index and q-deformed Yang-Mills},'' {\em JHEP} {\bf 10} (2013)
  018,
\href{http://www.arXiv.org/abs/1303.4460}{{\tt 1303.4460}}.

\bibitem{Bullimore:2014nla}
M.~Bullimore, M.~Fluder, L.~Hollands, and P.~Richmond, ``{The superconformal
  index and an elliptic algebra of surface defects},'' {\em JHEP} {\bf 10}
  (2014) 062,
\href{http://www.arXiv.org/abs/1401.3379}{{\tt 1401.3379}}.

\bibitem{Seiberg:1994rs}
N.~Seiberg and E.~Witten, ``{Electric - magnetic duality, monopole
  condensation, and confinement in N=2 supersymmetric Yang-Mills theory},''
  {\em Nucl. Phys.} {\bf B426} (1994) 19--52,
  \href{http://www.arXiv.org/abs/hep-th/9407087}{{\tt hep-th/9407087}}.
[Erratum: Nucl. Phys.B430,485(1994)].

\bibitem{Seiberg:1994aj}
N.~Seiberg and E.~Witten, ``{Monopoles, duality and chiral symmetry breaking in
  N=2 supersymmetric QCD},'' {\em Nucl. Phys.} {\bf B431} (1994) 484--550,
\href{http://www.arXiv.org/abs/hep-th/9408099}{{\tt hep-th/9408099}}.

\bibitem{Longhi:2012mj}
P.~Longhi, ``{The BPS Spectrum Generator In 2d-4d Systems},'' {\em JHEP} {\bf
  11} (2012) 107,
\href{http://www.arXiv.org/abs/1205.2512}{{\tt 1205.2512}}.

\bibitem{Gaiotto:2012rg}
D.~Gaiotto, G.~W. Moore, and A.~Neitzke, ``{Spectral networks},'' {\em Annales
  Henri Poincare} {\bf 14} (2013) 1643--1731,
\href{http://www.arXiv.org/abs/1204.4824}{{\tt 1204.4824}}.

\bibitem{Longhi:2016rjt}
P.~Longhi and C.~Y. Park, ``{ADE Spectral Networks},'' {\em JHEP} {\bf 08}
  (2016) 087,
\href{http://www.arXiv.org/abs/1601.02633}{{\tt 1601.02633}}.

\bibitem{Longhi:2016bte}
P.~Longhi and C.~Y. Park, ``{ADE Spectral Networks and Decoupling Limits of
  Surface Defects},'' {\em JHEP} {\bf 02} (2017) 011,
\href{http://www.arXiv.org/abs/1611.09409}{{\tt 1611.09409}}.

\bibitem{Cecotti:1992rm}
S.~Cecotti and C.~Vafa, ``{On classification of N=2 supersymmetric theories},''
  {\em Commun.Math.Phys.} {\bf 158} (1993) 569--644,
\href{http://www.arXiv.org/abs/hep-th/9211097}{{\tt hep-th/9211097}}.

\bibitem{Cordova:2015nma}
C.~C\'ordova and S.-H. Shao, ``{Schur Indices, BPS Particles, and
  Argyres-Douglas Theories},'' {\em JHEP} {\bf 01} (2016) 040,
\href{http://www.arXiv.org/abs/1506.00265}{{\tt 1506.00265}}.

\bibitem{Cecotti:2015lab}
S.~Cecotti, J.~Song, C.~Vafa, and W.~Yan, ``{Superconformal Index, BPS
  Monodromy and Chiral Algebras},''
\href{http://www.arXiv.org/abs/1511.01516}{{\tt 1511.01516}}.

\bibitem{Cordova:2016uwk}
C.~C\'ordova, D.~Gaiotto, and S.-H. Shao, ``{Infrared Computations of Defect
  Schur Indices},'' {\em JHEP} {\bf 11} (2016) 106,
\href{http://www.arXiv.org/abs/1606.08429}{{\tt 1606.08429}}.

\bibitem{Kinney:2005ej}
J.~Kinney, J.~M. Maldacena, S.~Minwalla, and S.~Raju, ``{An Index for 4
  dimensional super conformal theories},'' {\em Commun. Math. Phys.} {\bf 275}
  (2007) 209--254,
\href{http://www.arXiv.org/abs/hep-th/0510251}{{\tt hep-th/0510251}}.

\bibitem{Gadde:2011ik}
A.~Gadde, L.~Rastelli, S.~S. Razamat, and W.~Yan, ``{The 4d Superconformal
  Index from q-deformed 2d Yang-Mills},'' {\em Phys. Rev. Lett.} {\bf 106}
  (2011) 241602,
\href{http://www.arXiv.org/abs/1104.3850}{{\tt 1104.3850}}.

\bibitem{Gadde:2011uv}
A.~Gadde, L.~Rastelli, S.~S. Razamat, and W.~Yan, ``{Gauge Theories and
  Macdonald Polynomials},'' {\em Commun.Math.Phys.} {\bf 319} (2013) 147--193,
\href{http://www.arXiv.org/abs/1110.3740}{{\tt 1110.3740}}.

\bibitem{Gadde:2009kb}
A.~Gadde, E.~Pomoni, L.~Rastelli, and S.~S. Razamat, ``{S-duality and 2d
  Topological QFT},'' {\em JHEP} {\bf 1003} (2010) 032,
\href{http://www.arXiv.org/abs/0910.2225}{{\tt 0910.2225}}.

\bibitem{Beem:2013sza}
C.~Beem, M.~Lemos, P.~Liendo, W.~Peelaers, L.~Rastelli, {\em et al.},
  ``{Infinite Chiral Symmetry in Four Dimensions},'' {\em Commun.Math.Phys.}
  {\bf 336} (2015), no.~3, 1359--1433,
\href{http://www.arXiv.org/abs/1312.5344}{{\tt 1312.5344}}.

\bibitem{Iqbal:2012xm}
A.~Iqbal and C.~Vafa, ``{BPS Degeneracies and Superconformal Index in Diverse
  Dimensions},'' {\em Phys. Rev.} {\bf D90} (2014), no.~10, 105031,
\href{http://www.arXiv.org/abs/1210.3605}{{\tt 1210.3605}}.

\bibitem{DFZ}
T.~Dumitrescu, G.~Festuccia, and M.~Del~Zotto, ``{A Supersymmetric Index for
  Non-Conformal N=2 Theories in Four Dimensions},'' {\em to appear}.

\bibitem{Beem:2014rza}
C.~Beem, W.~Peelaers, L.~Rastelli, and B.~C. van Rees, ``{Chiral algebras of
  class S},'' {\em JHEP} {\bf 1505} (2015) 020,
\href{http://www.arXiv.org/abs/1408.6522}{{\tt 1408.6522}}.

\bibitem{Lemos:2014lua}
M.~Lemos and W.~Peelaers, ``{Chiral Algebras for Trinion Theories},'' {\em
  JHEP} {\bf 1502} (2015) 113,
\href{http://www.arXiv.org/abs/1411.3252}{{\tt 1411.3252}}.

\bibitem{Liendo:2015ofa}
P.~Liendo, I.~Ramirez, and J.~Seo, ``{Stress-tensor OPE in $ \mathcal{N}=2 $
  superconformal theories},'' {\em JHEP} {\bf 02} (2016) 019,
\href{http://www.arXiv.org/abs/1509.00033}{{\tt 1509.00033}}.

\bibitem{Lemos:2015orc}
M.~Lemos and P.~Liendo, ``{$\mathcal{N}=2$ central charge bounds from $2d$
  chiral algebras},'' {\em JHEP} {\bf 04} (2016) 004,
\href{http://www.arXiv.org/abs/1511.07449}{{\tt 1511.07449}}.

\bibitem{arakawa2015joseph}
T.~Arakawa and A.~Moreau, ``{Joseph ideals and lisse minimal W-algebras},''
\href{http://www.arXiv.org/abs/1506.00710}{{\tt 1506.00710}}.

\bibitem{Nishinaka:2016hbw}
T.~Nishinaka and Y.~Tachikawa, ``{On 4d rank-one N=3 superconformal field
  theories},''
\href{http://www.arXiv.org/abs/1602.01503}{{\tt 1602.01503}}.

\bibitem{Buican:2016arp}
M.~Buican and T.~Nishinaka, ``{Conformal Manifolds in Four Dimensions and
  Chiral Algebras},''
\href{http://www.arXiv.org/abs/1603.00887}{{\tt 1603.00887}}.

\bibitem{Arakawa:2016hkg}
T.~Arakawa and K.~Kawasetsu, ``{Quasi-lisse vertex algebras and modular linear
  differential equations},''
\href{http://www.arXiv.org/abs/1610.05865}{{\tt 1610.05865}}.

\bibitem{Bonetti:2016nma}
F.~Bonetti and L.~Rastelli, ``{Supersymmetric Localization in AdS$_5$ and the
  Protected Chiral Algebra},''
\href{http://www.arXiv.org/abs/1612.06514}{{\tt 1612.06514}}.

\bibitem{beem}
C.~Beem and L.~Rastelli, ``{Vertex Operators, Higgs Branches, and Modular
  Differential Equations},'' {\em to appear}.

\bibitem{CGS}
C.~C\'{o}rdova, D.~Gaiotto, and S.-H. Shao, ``{Surface Defects and Chiral
  Algebras},'' {\em to appear}.

\bibitem{BPR}
C.~Beem, W.~Peelaers, and L.~Rastelli, work in~progress.

\bibitem{Kontsevich:2008fj}
M.~Kontsevich and Y.~Soibelman, ``{Stability structures, motivic
  Donaldson-Thomas invariants and cluster transformations},''
\href{http://www.arXiv.org/abs/0811.2435}{{\tt 0811.2435}}.

\bibitem{Gaiotto:2008cd}
D.~Gaiotto, G.~W. Moore, and A.~Neitzke, ``{Four-dimensional wall-crossing via
  three-dimensional field theory},'' {\em Commun. Math. Phys.} {\bf 299} (2010)
  163--224,
\href{http://www.arXiv.org/abs/0807.4723}{{\tt 0807.4723}}.

\bibitem{Dimofte:2009tm}
T.~Dimofte, S.~Gukov, and Y.~Soibelman, ``{Quantum Wall Crossing in N=2 Gauge
  Theories},'' {\em Lett. Math. Phys.} {\bf 95} (2011) 1--25,
\href{http://www.arXiv.org/abs/0912.1346}{{\tt 0912.1346}}.

\bibitem{Dimofte:2009bv}
T.~Dimofte and S.~Gukov, ``{Refined, Motivic, and Quantum},'' {\em Lett. Math.
  Phys.} {\bf 91} (2010) 1,
\href{http://www.arXiv.org/abs/0904.1420}{{\tt 0904.1420}}.

\bibitem{Gaiotto:2009hg}
D.~Gaiotto, G.~W. Moore, and A.~Neitzke, ``{Wall-crossing, Hitchin Systems, and
  the WKB Approximation},''
\href{http://www.arXiv.org/abs/0907.3987}{{\tt 0907.3987}}.

\bibitem{Galakhov:2014xba}
D.~Galakhov, P.~Longhi, and G.~W. Moore, ``{Spectral Networks with Spin},''
  {\em Commun. Math. Phys.} {\bf 340} (2015), no.~1, 171--232,
\href{http://www.arXiv.org/abs/1408.0207}{{\tt 1408.0207}}.

\bibitem{Gaiotto:2010be}
D.~Gaiotto, G.~W. Moore, and A.~Neitzke, ``{Framed BPS States},'' {\em Adv.
  Theor. Math. Phys.} {\bf 17} (2013), no.~2, 241--397,
\href{http://www.arXiv.org/abs/1006.0146}{{\tt 1006.0146}}.

\bibitem{Cecotti:2010fi}
S.~Cecotti, A.~Neitzke, and C.~Vafa, ``{R-Twisting and 4d/2d
  Correspondences},''
\href{http://www.arXiv.org/abs/1006.3435}{{\tt 1006.3435}}.

\bibitem{Gaiotto:2015zna}
D.~Gaiotto, G.~W. Moore, and E.~Witten, ``{An Introduction To The Web-Based
  Formalism},''
\href{http://www.arXiv.org/abs/1506.04086}{{\tt 1506.04086}}.

\bibitem{Gaiotto:2015aoa}
D.~Gaiotto, G.~W. Moore, and E.~Witten, ``{Algebra of the Infrared: String
  Field Theoretic Structures in Massive ${\cal N}=(2,2)$ Field Theory In Two
  Dimensions},''
\href{http://www.arXiv.org/abs/1506.04087}{{\tt 1506.04087}}.

\bibitem{Lee:2011ph}
S.~Lee and P.~Yi, ``{Framed BPS States, Moduli Dynamics, and Wall-Crossing},''
  {\em JHEP} {\bf 04} (2011) 098,
\href{http://www.arXiv.org/abs/1102.1729}{{\tt 1102.1729}}.

\bibitem{Cordova:2013bza}
C.~C\'ordova and A.~Neitzke, ``{Line Defects, Tropicalization, and
  Multi-Centered Quiver Quantum Mechanics},'' {\em JHEP} {\bf 09} (2014) 099,
\href{http://www.arXiv.org/abs/1308.6829}{{\tt 1308.6829}}.

\bibitem{Moore:2015szp}
G.~W. Moore, A.~B. Royston, and D.~V.~d. Bleeken, ``{Semiclassical framed BPS
  states},''
\href{http://www.arXiv.org/abs/1512.08924}{{\tt 1512.08924}}.

\bibitem{Gabella:2016zxu}
M.~Gabella, ``{Quantum Holonomies from Spectral Networks and Framed BPS
  States},''
\href{http://www.arXiv.org/abs/1603.05258}{{\tt 1603.05258}}.

\bibitem{Brennan:2016znk}
T.~D. Brennan and G.~W. Moore, ``{A note on the semiclassical formulation of
  BPS states in four-dimensional $N=$ 2 theories},'' {\em PTEP} {\bf 2016}
  (2016), no.~12, 12C110,
\href{http://www.arXiv.org/abs/1610.00697}{{\tt 1610.00697}}.

\bibitem{Benini:2013nda}
F.~Benini, R.~Eager, K.~Hori, and Y.~Tachikawa, ``{Elliptic genera of
  two-dimensional N=2 gauge theories with rank-one gauge groups},'' {\em Lett.
  Math. Phys.} {\bf 104} (2014) 465--493,
\href{http://www.arXiv.org/abs/1305.0533}{{\tt 1305.0533}}.

\bibitem{Benini:2013xpa}
F.~Benini, R.~Eager, K.~Hori, and Y.~Tachikawa, ``{Elliptic Genera of 2d
  ${\mathcal{N}}$ = 2 Gauge Theories},'' {\em Commun. Math. Phys.} {\bf 333}
  (2015), no.~3, 1241--1286,
\href{http://www.arXiv.org/abs/1308.4896}{{\tt 1308.4896}}.

\bibitem{Witten:1986bf}
E.~Witten, ``{Elliptic Genera and Quantum Field Theory},'' {\em Commun. Math.
  Phys.} {\bf 109} (1987)
525.

\bibitem{Bershadsky:1993cx}
M.~Bershadsky, S.~Cecotti, H.~Ooguri, and C.~Vafa, ``{Kodaira-Spencer theory of
  gravity and exact results for quantum string amplitudes},'' {\em Commun.
  Math. Phys.} {\bf 165} (1994) 311--428,
\href{http://www.arXiv.org/abs/hep-th/9309140}{{\tt hep-th/9309140}}.

\bibitem{Cecotti:1992qh}
S.~Cecotti, P.~Fendley, K.~A. Intriligator, and C.~Vafa, ``{A New
  supersymmetric index},'' {\em Nucl. Phys.} {\bf B386} (1992) 405--452,
\href{http://www.arXiv.org/abs/hep-th/9204102}{{\tt hep-th/9204102}}.

\bibitem{Dorey:1998yh}
N.~Dorey, ``{The BPS spectra of two-dimensional supersymmetric gauge theories
  with twisted mass terms},'' {\em JHEP} {\bf 11} (1998) 005,
\href{http://www.arXiv.org/abs/hep-th/9806056}{{\tt hep-th/9806056}}.

\bibitem{CD}
C.~C\'{o}rdova and T.~Dumitrescu, ``{Current Algebra Constraints on BPS
  Particles},'' {\em to appear}.

\bibitem{Witten:1979ey}
E.~Witten, ``{Dyons of Charge e theta/2 pi},'' {\em Phys. Lett.} {\bf B86}
  (1979)
283--287.

\end{thebibliography}\endgroup
\bibliographystyle{utphys}

\end{document}